\documentclass[prc,aps,manuscript]{revtex4}
\usepackage{graphicx}

\begin{document}
\draft
\title{Shape fluctuations in a Fermi system with nonlinear dissipativity}
\author{V.M. Kolomietz$^{1,2)}$, S.V. Lukyanov$^{1)}$ and S. Shlomo$^{2)}$}
\address{$^{1)}$Institute for Nuclear Research, Prosp. Nauki 47, 03680 Kiev, \\
Ukraine}
\address{$^{2)}$Cyclotron Institute, Texas A\&M University, College Station, \\
Texas 77843}


\begin{abstract}
The contribution of thermal fluctuations to the widths of isoscalar giant
multipole resonances (GMR) in heated nuclei is studied. Starting from the
collisional kinetic equation, it is shown that an additional contribution to
the nuclear friction and the corresponding GMR widths arises due to the
nonlinear dissipativity effect. It is also shown that the magnitude of the
contributions of the thermal fluctuations to the nuclear friction
coefficient and the GMR widths do not exceed $\sim $ 20\%.

\bigskip


\end{abstract}



\maketitle

\section{Introduction}

In general, the damping of collective excitations in a cold finite Fermi
system, e.g. the width of a giant multipole resonance (GMR) in cold nuclei,
is determined by the two-body collisions, the particle-hole energy
fragmentation (Landau damping) and the escape width. The damping in cold
nuclei was intensively investigated in both the quantum (RPA like) \cite
{reya,wamb,brva1,brva2,yan2} and the semiclassical (kinetic theory)
approaches \cite{bert,bbb2,cadi,smbo,aybo,koma1,hass}. The collisional
damping is due to the coupling of particle-hole excitations to more
complicated states. In the kinetic theory, this type of damping is simulated
by the collision integral and leads to a collisional component of the
intrinsic width of the collective eigenstates. The fragmentation width is
caused by the interaction of particles with the time-dependent
self-consistent mean field. This contribution to the intrinsic width does
not reflect a motion of the system toward the thermal equilibrium but
indicates rather a redistribution of the particle-hole excitations in the
vicinity of the collective state. In a hot system an additional contribution
to the damping of the collective excitations arises through thermodynamic
fluctuations of the corresponding collective variables because of the
fluctuation-dissipation theorem. In this context, one of the most important
open problem is the behavior of the GMR width in hot nuclei as a function of
the temperature $T$. There are two essential different sources for the $T$%
-dependence of the GMR width. The first one is given by the thermal
contribution to the damping width from an increasing nucleon-nucleon
collision rate ($2p2h$ excitations) plus a Landau spreading due to thermally
allowed $ph$ transitions \cite{brva1,brva2,vinh,kopl1,kola,diko}. In the
second one a temperature increasing of the width\ is caused by the coupling
of the GDR to the thermal shape fluctuations of the nucleus \cite
{orca,blca,orbo}. In the present work, we study a new effect of the
influence of the thermal shape fluctuations of the nucleus on the damping of
the collective motion caused by the \underline{nonlinear dissipativity}
appearing in the higher order variations of the collision integral. We point
out that in the commonly used linear order of the variation of the collision
integral, the thermal fluctuations do not lead to dissipation (viscosity) in
the macroscopic equations of motion because the following ensemble smearing
of the kinetic equation washes out the fluctuation terms from the final
macroscopic equations of motion. This paper is organized as follows. In
section II we suggest a proof of the Langevin equation for nuclear local
variables (particle density, velocity field and pressure tensor), starting
from the collisional kinetic equation. We perform a high order expansion of
the collisional integral and derive the non-Markovian pressure tensor to the
Navier--Stokes-like equation of motion. In section III we carry out the
ensemble averaging and reduce the local equations of motion to the
macroscopic form and derive the macroscopic response function. In the
derivations, the main features of the dynamical distortion of the Fermi
surface are taken into account. Results of numerical calculations are
presented in section IV. We conclude and summarize in section V. The
Appendixes provide a derivation of the high order variations of the
collision integral with respect to the variation of the phase space
distribution function.

\section{Local equations of motion}

We will start from the collisional kinetic equation in the following form
\cite{AbKh.RPP.59}
\begin{equation}
\frac{\partial f}{\partial t}+\frac{\vec{p}}{m}\frac{\partial f}{\partial
\vec{r}}-\frac{\partial V}{\partial \vec{r}}\frac{\partial f}{\partial \vec{p%
}}\ +\ \stackrel{\rightarrow }{F}_{{\rm ext}}\frac{\partial f}{\partial \vec{%
p}}=\delta {\rm St}[f]+y.  \label{lv}
\end{equation}
Here, $f\equiv f(\vec{r},\vec{p},t)$ is the Wigner distribution function, $%
V\equiv V(\vec{r},\vec{p},t)$ is the self-consistent mean field, $\stackrel{%
\rightarrow }{F}_{{\rm ext}}$ is the external driving force and $\delta {\rm %
St}[f]$ is the collision integral which takes into account the memory
effects \cite{KoPl.YaF.94}. The random variable $y\equiv y(\vec{r},\vec{p},t)
$ in Eq. (\ref{lv}) represents the random force. As such, its ensemble
average vanishes, $\left\langle y\right\rangle =0$. To reduce Eq. (\ref{lv})
to closed equations of motion for the macroscopic collective variables we
will follow the nuclear fluid dynamic approach \cite{ber1,he1,ns,wa,kol1}
and take into account the dynamic Fermi-surface distortion up to
multipolarity $l=2$. Evaluating the first three moments of Eq. (\ref{lv}) in
the ${\vec{p}}$ space, we reduce Eq. (\ref{lv}) to the hydrodynamic-like
equations of motion for particle density $\rho ,$\ velocity field $\stackrel{%
\rightarrow }{u}$\ and the pressure tensor $P_{\alpha \beta }$\ (for
details, see Ref. \cite{KoTa.PS.81})
\begin{equation}
{\frac{\partial \rho }{\partial t}}=-{\frac{\partial }{\partial r_{\nu }}}%
\rho u_{\nu },  \label{sys1}
\end{equation}
\begin{equation}
{\frac{\partial }{\partial t}}m\rho u_{\alpha }+{\frac{\partial }{\partial
r_{\nu }}}m\rho u_{\nu }u_{\alpha }+{\frac{\partial }{\partial r_{\nu }}}%
P_{\nu \alpha }+\rho {\frac{\partial }{\partial r}}_{\alpha }V-\rho F_{{\rm %
ext},\alpha }=0,  \label{sys2}
\end{equation}
and
\begin{equation}
{\frac{\partial }{\partial t}}P_{\alpha \beta }+{\frac{\partial }{\partial r}%
}_{\nu }u_{\nu }P_{\alpha \beta }+P_{\nu \beta }{\frac{\partial }{\partial r}%
}_{\nu }u_{\alpha }+P_{\nu \alpha }{\frac{\partial }{\partial r}}_{\nu
}u_{\beta }=Q_{\alpha \beta }+y_{\alpha \beta }.  \label{sys3}
\end{equation}
Here,
\begin{equation}
\rho =\int {\frac{gd{\vec{p}}}{(2\pi \hbar )^{3}}}f,\quad {\vec{u}}={\frac{1%
}{\rho }}\int {\frac{gd{\vec{p}}}{(2\pi \hbar )^{3}}}{\frac{{\vec{p}}}{m}}%
f,\quad P_{\alpha \beta }={\frac{1}{m}}\int {\frac{gd{\vec{p}}}{(2\pi \hbar
)^{3}}}(p_{\alpha }-mu_{\alpha })(p_{\beta }-mu_{\beta })f,  \label{deriv1}
\end{equation}
$g=4$ is the spin-isospin degeneracy factor, $Q_{\alpha \beta }$\ is
associated with dissipative processes
\begin{equation}
Q_{\alpha \beta }={\frac{1}{m}}\int {\frac{gd{\vec{p}}}{(2\pi \hbar )^{3}}}%
(p_{\alpha }-mu_{\alpha })(p_{\beta }-mu_{\beta })\delta {\rm St}[f],
\label{qab}
\end{equation}
and $y_{\alpha \beta }$\ gives the contribution from the random force
\begin{equation}
y_{\alpha \beta }={\frac{1}{m}}\int {\frac{gd{\vec{p}}}{(2\pi \hbar )^{3}}\ }%
p_{\alpha }p_{\beta }\ y.  \label{rand1}
\end{equation}
In Eqs. (\ref{sys1}) to (\ref{sys3}) and in the following expressions,
repeated Greek indices are to be understood as summed over. The pressure
tensor $P_{\alpha \beta }$ can be written as
\begin{equation}
P_{\alpha \beta }=P_{{\rm eq}}\delta _{\alpha \beta }+P_{\alpha \beta
}^{\prime },  \label{p}
\end{equation}
where
\begin{equation}
P_{\alpha \beta }^{\prime }={\frac{1}{m}}\int {\frac{gd{\vec{p}}}{(2\pi
\hbar )^{3}}}(p_{\alpha }-mu_{\alpha })(p_{\beta }-mu_{\beta })\delta f_{2},
\label{dpi}
\end{equation}
$\delta f_{2}\equiv \delta f_{2}({\bf r},{\bf p};t)$ is the small quadrupole
deviation of the distribution function $f$ from the one in equilibrium, $f_{%
{\rm eq}}\equiv f_{{\rm eq}}({\vec{r}},{\vec{p}})$, and $P_{{\rm eq}}$ is
the pressure due to the Fermi motion of the nucleons in the ground state of
the nucleus. Assuming the Thomas-Fermi approximation for $f_{{\rm eq}}$, one
has
\[
P_{{\rm eq}}=\frac{2}{3}\ \int {\frac{gd{\vec{p}}}{(2\pi \hbar )^{3}}}\,{%
\frac{p^{2}}{2m}}f_{{\rm eq}}={\frac{2}{5}}{\frac{\hbar ^{2}}{m}}\left( {{%
\frac{6{\pi }^{2}}{g}}}\right) ^{2/3}\rho _{{\rm eq}}^{5/3},
\]
where $\rho _{{\rm eq}}\equiv \rho _{{\rm eq}}({\vec{r}})$ is the particle
density in equilibrium. Using Eq. (\ref{p}) we will rewrite Eq. (\ref{sys3})
as
\begin{equation}
{\frac{\partial }{\partial t}}P_{\alpha \beta }^{\prime }+P_{{\rm eq}%
}\Lambda _{\alpha \beta }+\hat{L}P_{\alpha \beta }^{\prime }=Q_{\alpha \beta
}+y_{\alpha \beta },  \label{sys3.1}
\end{equation}
where
\begin{equation}
\Lambda _{\alpha \beta }={\frac{\partial u_{\alpha }}{\partial r_{\beta }}}+{%
\frac{\partial u_{\beta }}{\partial r_{\alpha }}}-{\frac{2}{3}}\delta
_{\alpha \beta }\frac{\partial u_{\nu }}{\partial r_{\nu }},
\end{equation}
and the operator $\hat{L}$ is defined by
\begin{equation}
\hat{L}P_{\alpha \beta }^{\prime }=u_{\nu }{\frac{\partial }{\partial r_{\nu
}}}P_{\alpha \beta }^{\prime }+P_{\alpha \beta }^{\prime }\frac{\partial
u_{\nu }}{\partial r_{\nu }}+P_{\alpha \nu }^{\prime }{\frac{\partial
u_{\beta }}{\partial r_{\nu }}}+P_{\beta \nu }^{\prime }{\frac{\partial
u_{\alpha }}{\partial r_{\nu }}}.
\end{equation}
To evaluate the tensor $Q_{\alpha \beta }$\ in Eq. (\ref{sys3.1}) we will
use the collision integral in the following general form \cite{AbKh.RPP.59}
\begin{equation}
\delta {\rm St}[f]=\int {\frac{g^{2}d\vec{p}_{2}d\vec{p}_{3}d\vec{p}_{4}}{%
(2\pi \hbar )^{6}}}w(\{\vec{p}_{j}\})Q(\{f_{j}\})\ \delta (\Delta \epsilon
)\delta (\Delta \vec{p}),  \label{int1}
\end{equation}
where $w(\{\vec{p}_{j}\})\equiv w(\vec{p}_{1},\vec{p}_{2};\vec{p}_{3},\vec{p}%
_{4})$ is the spin-isospin averaged probability for two-body scattering, $%
Q(\{f_{j}\})=f_{1}f_{2}(1-f_{3})(1-f_{4})-(1-f_{1})(1-f_{2})f_{3}f_{4}$ is
the Pauli blocking factor, $\Delta \vec{p}=\vec{p}_{1}+\vec{p}_{2}-\vec{p}%
_{3}-\vec{p}_{4}$ and $\Delta \epsilon =\epsilon _{1}+\epsilon _{2}-\epsilon
_{3}-\epsilon _{4}$, with $\epsilon _{j}=p_{j}^{2}/2m+V(r_{j})$ being the
single-particle energy. In performing the variation of the Pauli blocking
factor $Q(\{f_{j}\})$\ with respect to $\delta f$\ ,\ we will keep all the
terms, up to the third order in $\delta f$. The collision integral then
takes the following form (see Appendix A)
\begin{equation}
\delta {\rm St=\ }\delta {\rm St}_{1}+\delta {\rm St}_{2}+\delta {\rm St}%
_{3},  \label{int2}
\end{equation}
where $\delta {\rm St}_{n}$ is the variation of the collision integral $%
\delta {\rm St}$\ in the $n$-th order of $\ \delta f$. Considering Eqs. (\ref
{qab}) and (\ref{int2}), the tensor $Q_{\alpha \beta }$\ can be written as
\begin{equation}
Q_{\alpha \beta }=Q_{\alpha \beta }^{(1)}+Q_{\alpha \beta }^{(2)}+Q_{\alpha
\beta }^{(3)},  \label{qq1q2q3}
\end{equation}
where $Q_{\alpha \beta }^{(n)}$\ is due to the contribution of $\delta {\rm %
St}_{n}$ in Eq. (\ref{qab}). The first order term $Q_{\alpha \beta }^{(1)}$\
is simplified as \cite{KoMaPl.NP.92}
\begin{equation}
Q_{\alpha \beta }^{(1)}=-{\frac{1}{\tau _{2}}}P_{\alpha \beta }^{\prime },
\label{q1}
\end{equation}
where $\tau _{2}$ is the two-body relaxation time in the case of quadrupole
deformation of the Fermi surface. The higher order terms $Q_{\alpha \beta
}^{(2)}$ and$\ Q_{\alpha \beta }^{(3)}$\ can be reduced to the following
forms (see Appendix A, Eqs. (\ref{q7y}) and (\ref{q7x}))
\begin{equation}
Q_{\alpha \beta }^{(2)}={\frac{mP_{0}^{\prime }}{\zeta }\ }P_{\alpha \beta
}^{\prime },\quad Q_{\alpha \beta }^{(3)}={\frac{m^{2}P_{0}^{\prime 2}}{\xi }%
}P_{\alpha \beta }^{\prime },  \label{q23}
\end{equation}
where the quantities $\zeta $ and $\xi $ are deduced from the collision
integral (see Appendix A, Eqs. (\ref{zetaend}) and (\ref{xiend})) and
\begin{equation}
P_{0}^{\prime }=\frac{1}{2}(P_{xx}^{\prime }+P_{yy}^{\prime }-P_{zz}^{\prime
}).  \label{p0}
\end{equation}
To simplify the derivations, we will introduce the operator $\hat{N}$\ as \
\
\begin{equation}
\hat{N}P_{\alpha \beta }^{\prime }=Q_{\alpha \beta }^{(2)}+Q_{\alpha \beta
}^{(3)}.  \label{opn}
\end{equation}
Equation (\ref{sys3.1}) is then rewritten as
\begin{equation}
{\frac{\partial }{\partial t}}P_{\alpha \beta }^{\prime }+P_{{\rm eq}%
}\Lambda _{\alpha \beta }+{\frac{1}{\tau _{2}}}P_{\alpha \beta }^{\prime }+%
\hat{L}P_{\alpha \beta }^{\prime }=\hat{N}P_{\alpha \beta }^{\prime
}+y_{\alpha \beta }.  \label{sys3.2}
\end{equation}
We will look for a solution of Eq. (\ref{sys3.2}) in the following form
\begin{equation}
P_{\alpha \beta }^{\prime }=P_{\alpha \beta }^{\prime (0)}+P_{\alpha \beta
}^{\prime (NL)}.  \label{p0n}
\end{equation}
Here, the tensor $P_{\alpha \beta }^{\prime (0)}$\ is obtained as a solution
to the following linear differential equation \cite{Ko.b.90}
\begin{equation}
{\frac{\partial }{\partial t}}P_{\alpha \beta }^{\prime (0)}+P_{{\rm eq}%
}\Lambda _{\alpha \beta }+{\frac{1}{\tau _{2}}}P_{\alpha \beta }^{\prime
(0)}=y_{\alpha \beta },  \label{sys3.3}
\end{equation}
and it is given by the following non-Markovian form
\begin{equation}
P_{\alpha \beta }^{\prime (0)}(t)=-\int_{-\infty }^{t}dt^{\prime }\ \exp
\left( \frac{t^{\prime }-t}{\tau _{2}}\right) \left[ P_{{\rm eq}}\ \Lambda
_{\alpha \beta }(t^{\prime })-y_{\alpha \beta }(t^{\prime })\right] .
\label{pi0}
\end{equation}
The tensor $P_{\alpha \beta }^{\prime (NL)}$ in Eq. (\ref{p0n}) satisfies
the nonlinear\ differential equation
\begin{equation}
{\frac{\partial }{\partial t}}P_{\alpha \beta }^{\prime (NL)}+{\frac{1}{\tau
_{2}}}P_{\alpha \beta }^{\prime (NL)}+\hat{L}P_{\alpha \beta }^{\prime }-%
\hat{N}P_{\alpha \beta }^{\prime }=0.  \label{sys3.4}
\end{equation}
To solve Eq. (\ref{sys3.4}), we will use the iteration procedure. The first
order iteration $P_{\alpha \beta ,1}^{\prime (NL)}(t^{\prime })$ to Eq. (\ref
{sys3.4}) reads
\begin{equation}
P_{\alpha \beta ,1}^{\prime (NL)}(t)=\int_{-\infty }^{t}dt^{\prime }\ \exp
\left( \frac{t^{\prime }-t}{\tau _{2}}\right) \ \left[ \hat{N}P_{\alpha
\beta }^{\prime (0)}(t^{\prime })-\hat{L}P_{\alpha \beta }^{\prime
(0)}(t^{\prime })\right] ,  \label{pin1}
\end{equation}
and the second iteration is given by
\[
P_{\alpha \beta ,2}^{\prime (NL)}(t)=\int_{-\infty }^{t}dt^{\prime }\ \exp
\left( \frac{t^{\prime }-t}{\tau _{2}}\right) \
\]
\begin{equation}
\times \left[ \hat{N}\left( P_{\alpha \beta }^{\prime (0)}(t^{\prime
})+P_{\alpha \beta ,1}^{\prime (NL)}(t^{\prime })\right) -\hat{L}\left(
P_{\alpha \beta }^{\prime (0)}(t^{\prime })+P_{\alpha \beta ,1}^{\prime
(NL)}(t^{\prime })\right) \right] .  \label{pin}
\end{equation}
Below, we will apply Eqs. (\ref{sys1}) to (\ref{sys3}) to the
small-amplitude vibrations of the particle density $\delta \rho $ near the
equilibrium$.$ We point out that we do not assume the velocity field $\vec{u}
$ to be small. Finally, taking into account the above mentioned derivations
we will rewrite Eq. (\ref{sys2}) as
\begin{equation}
m\rho _{{\rm eq}}{\frac{\partial u_{\alpha }}{\partial t}}+m\rho _{{\rm eq}%
}u_{\nu }{\frac{\partial u_{\alpha }}{\partial r_{\nu }}}+\rho _{{\rm eq}}{%
\frac{\partial }{\partial r_{\alpha }}}\left[ \left. {\frac{\delta
^{2}\varepsilon }{\delta \rho ^{2}}}\right| _{{\rm eq}}\delta \rho \right] +{%
\frac{\partial P_{\alpha \beta }^{\prime (0)}}{\partial r_{\alpha }}}+{\frac{%
\partial P_{\alpha \beta }^{\prime (NL)}}{\partial r_{\alpha }}}-\rho _{{\rm %
eq}}F_{{\rm ext},\alpha }=0  \label{sys2.2}
\end{equation}
where $\varepsilon $ is the particle energy density. \bigskip

\section{Ensemble averaging and macroscopic response}

Let us introduce the displacement field $\vec{\chi}$ related to the velocity
field $\vec{u}$ by $\vec{u}(\vec{r},t)=\dot{\vec{\chi}}(\vec{r},t)$, where
the dot denotes a time derivative. For the displacement field we will assume
the following separable form $\vec{\chi}(\vec{r},t)=\beta (t)\vec{v}(\vec{r})
$. Using this separable form of $\vec{\chi}(\vec{r},t)$, we reduce Eq. (\ref
{sys2.2}) to the equation of motion for the collective variable $\beta (t)$
in the presence of the external field ${\cal F}_{{\rm ext}}(t)$ and the
random force $\widetilde{y}(t)$ (see {\rm Appendix C}, Eq. (\ref{eqbeta})).
Below, we will look for the response of a nucleus to the periodic external
field ${\cal F}_{{\rm ext}}(t)={\cal F}_{\omega }\exp (i\omega t)$.\ Because
of the random force $\widetilde{y}(t)$\ in Eq. (\ref{eqbeta}), we will
separate the description of the collective motion into two parts with $\beta
(t)=\widetilde{\beta }(t)+\delta \beta (t)$. The first motion is related to
the driving force ${\cal F}_{{\rm ext}}(t)$ and it is associated with the
velocity $\dot{\widetilde{\beta }}$. The second one is due to the random
force $\widetilde{y}(t)$\ with the velocity $\delta \dot{\beta}$. We will
assume that $|\delta \dot{\beta}|\gg |\dot{\widetilde{\beta }}|.$ Performing
the ensemble averaging, one can write
\[
\left\langle \dot{\beta}(t_{1})\dot{\beta}(t_{2})\dot{\beta}%
(t_{3})\right\rangle \approx \ \dot{\widetilde{\beta }}(t_{1})\left\langle
\delta \dot{\beta}(t_{2})\ \delta \dot{\beta}(t_{3})\right\rangle +\dot{%
\widetilde{\beta }}(t_{2})\left\langle \delta \dot{\beta}(t_{1})\ \delta
\dot{\beta}(t_{3})\right\rangle
\]
\begin{equation}
+\dot{\widetilde{\beta }}(t_{3})\left\langle \delta \dot{\beta}(t_{2})\
\delta \dot{\beta}(t_{1})\right\rangle .  \label{split}
\end{equation}
We will also assume the following ergodic property for the correlation
function $\left\langle \delta \dot{\beta}(t)\ \delta \dot{\beta}(t^{\prime
})\right\rangle ,$ see\ Ref. \cite{LaLi.b.v5.76}, Ch.12,
\begin{equation}
\left\langle \delta \dot{\beta}(t)\ \delta \dot{\beta}(t^{\prime
})\right\rangle =\int_{-\infty }^{\infty }\frac{d\omega }{2\pi }\left(
\delta \dot{\beta}^{2}\right) _{\omega }e^{-i\omega (t-t^{\prime })}.
\label{erg}
\end{equation}
The Fourier component, $\left( \delta \dot{\beta}^{2}\right) _{\omega },$\
of the correlation function is governed by the correlation properties of the
random force $\widetilde{y}(t),$\ see below. The macroscopic equation of
motion (\ref{eqbeta}) is significantly simplified in the case of a Fermi
distribution for the equilibrium distribution function $f_{{\rm eq}}$%
\begin{equation}
f_{{\rm eq}}=\left[ 1+\exp \left( \frac{\epsilon -\epsilon _{F}}{T}\right) %
\right] ^{-1},  \label{f.fermi}
\end{equation}
where $\epsilon _{F}$\ is the Fermi energy. In this case, one obtains from
Eqs. (\ref{zetaend}) that $1/\zeta \ll 1$ (see also Figs. 1 and 2) and the
contribution of the terms with $A_{2},$ $A_{3}$ and $A_{5}$ in Eq. (\ref
{eqbeta}) is negligible. Performing the ensemble averaging of Eq. (\ref
{eqbeta}), using Eq. (\ref{split}) and $\left\langle y(t)\right\rangle =0$\
and assuming $\left\langle \beta (t)\right\rangle =\widetilde{\beta }(t)=%
\widetilde{\beta }_{\omega }\exp (i\omega t),$\ we reduce Eq. (\ref{eqbeta})
to the following form
\[
-B\omega {}^{2}\widetilde{\beta }_{\omega }+C_{LDM}\widetilde{\beta }%
_{\omega }+\frac{i\omega \tau _{2}}{1+i\omega \tau _{2}}\left( A_{0}+\tau
_{2}^{2}A_{4}\int_{-\infty }^{\infty }\frac{d\omega ^{\prime }}{2\pi }\frac{%
\left( \delta \dot{\beta}^{2}\right) _{\omega ^{\prime }}}{1+i\omega
^{\prime }\tau _{2}}+\right. \frac{4\tau _{2}^{3}A_{1}}{1+i\omega \tau _{2}}%
\int_{-\infty }^{\infty }\frac{d\omega ^{\prime }}{2\pi }\frac{\left( \delta
\dot{\beta}^{2}\right) _{\omega ^{\prime }}}{1+(\omega ^{\prime }\tau
_{2})^{2}}
\]
\begin{equation}
+\tau _{2}^{2}A_{4}\int_{-\infty }^{\infty }\frac{d\omega ^{\prime }}{2\pi }%
\frac{\left( \delta \dot{\beta}^{2}\right) _{\omega ^{\prime }}}{1+i\omega
^{\prime }\tau _{2}}\frac{1}{1+i(\omega ^{\prime }+\omega )\tau _{2}}+\left.
\frac{\tau _{2}^{2}A_{4}}{1+i\omega \tau _{2}}\int_{-\infty }^{\infty }\frac{%
d\omega ^{\prime }}{2\pi }\frac{\left( \delta \dot{\beta}^{2}\right)
_{\omega ^{\prime }}}{1+i(\omega ^{\prime }+\omega )\tau _{2}}\right)
\widetilde{\beta }_{\omega }=B{\cal F}_{\omega }.  \label{dispend}
\end{equation}
Considering the nuclear isoscalar quadrupole mode, we will assume an
irrotational motion with the displacement field $\vec{v}(\stackrel{%
\rightarrow }{r})$\ given by \cite{kidks}
\begin{equation}
\vec{v}(\stackrel{\rightarrow }{r})=\stackrel{\rightarrow }{\nabla }%
(r^{2}Y_{20}(\hat{r}))/2,  \label{v1}
\end{equation}
and the time dependent radius of the nucleus given by
\begin{equation}
R(t)=R_{0}\left[ 1+\widetilde{\beta }(t)Y_{20}(\hat{r})\right] .  \label{rho}
\end{equation}
In this particular case the calculation of the coefficients $A_{1}$\ and $%
A_{4}$\ from Eqs.\ (\ref{a1}) and (\ref{a4}) gives: $A_{1}=16A_{0}(mP_{{\rm %
eq}})^{2}/\xi $ and $A_{4}=12A_{0}$. The mass coefficient $B$\ of Eq. (\ref
{b}) for the displacement field of Eq. (\ref{v1}) is given by
\begin{equation}
B=\frac{3}{8\pi }AmR_{0}^{2},  \label{b2}
\end{equation}
where $A$ is the nuclear mass number. Let us introduce the collective
response function $\chi (\omega )$\ as
\begin{equation}
\widetilde{\beta }_{\omega }=\chi (\omega )\ {\cal F}_{\omega }.  \label{chi}
\end{equation}
Using Eqs. (\ref{dispend}) and (\ref{rho}), we obtain from Eq. (\ref{chi})
\begin{equation}
\chi ^{-1}(\omega )=-\omega {}^{2}+\omega _{0}^{2}+i\omega \ \gamma
_{0}+12\tau _{2}^{2}\frac{A_{0}}{B}\frac{i\omega \tau _{2}}{1+i\omega \tau
_{2}}K(\omega ).  \label{chi-1}
\end{equation}
Here,
\begin{equation}
\omega _{0}=\sqrt{\frac{C_{LDM}+C^{\prime }(\omega )}{B}},\hspace{1cm}\gamma
_{0}=\frac{A_{0}}{B}\frac{\tau _{2}}{1+(\omega \tau _{2})^{2}}  \label{gam0}
\end{equation}
and
\[
K(\omega )=\int_{-\infty }^{\infty }\frac{d\omega ^{\prime }}{2\pi }\frac{%
\left( \delta \dot{\beta}^{2}\right) _{\omega ^{\prime }}}{1+i\omega
^{\prime }\tau _{2}}\left( 1+\frac{1}{1+i(\omega ^{\prime }+\omega )\tau _{2}%
}+\frac{4}{1+i\omega \tau _{2}}\frac{\tau _{2}/\tau ^{\prime \prime }}{%
1-i\omega ^{\prime }\tau _{2}}\right.
\]
\begin{equation}
+\left. \frac{1}{1+i\omega \tau _{2}}\frac{1+i\omega ^{\prime }\tau _{2}}{%
1+i(\omega ^{\prime }+\omega )\tau _{2}}\right) .  \label{s}
\end{equation}
We have also used the following notations
\begin{equation}
C^{\prime }(\omega )=A_{0}\frac{(\omega \tau _{2})^{2}}{1+(\omega \tau
_{2})^{2}}  \label{cs}
\end{equation}
and
\[
\tau ^{\prime \prime }=\xi /(mP_{{\rm eq}})^{2}.
\]
We point out that the additional contribution to the stiffness coefficient $%
C^{\prime }(\omega )$ in Eq. (\ref{gam0}) is caused by the distortion of the
Fermi surface \cite{Ko.b.90}. The expression (\ref{chi-1}) can be rewritten
as
\begin{equation}
\chi ^{-1}(\omega )=(\omega _{0}^{2}+\Delta \omega _{0}^{2}-\omega
^{2})+i\omega \ (\gamma _{0}+\Delta \gamma ),  \label{chi4}
\end{equation}
where we have introduced the following notations for the additional
components of the relaxation coefficient and the squared frequency $\omega $
\begin{equation}
\frac{\Delta \gamma }{\gamma _{0}}=12\tau _{2}^{2}\left\{ {\rm Re}K(\omega
)+\omega \tau _{2}{\rm Im}K(\omega )\right\} ,  \label{dgg0}
\end{equation}
\begin{equation}
\frac{\Delta \omega _{0}^{2}}{\omega _{0}^{2}}=12\tau _{2}^{2}\left\{ {\rm Re%
}K(\omega )-\frac{1}{\omega \tau _{2}}{\rm Im}K(\omega )\right\} \left( 1+%
\frac{C_{LDM}}{C^{\prime }(\omega )}\right) ^{-1}.  \label{dww0}
\end{equation}
Note that above, $\omega $ is real. Finally, the macroscopic strength
function $S(\omega )=-{\rm Im}\chi (\omega )$\ is given by
\begin{equation}
S(\omega )=\frac{(\gamma _{0}+\Delta \gamma )\omega }{(\omega
_{0}^{2}+\Delta \omega _{0}^{2}-\omega ^{2})^{2}+(\gamma _{0}+\Delta \gamma
)^{2}\omega ^{2}}.  \label{str}
\end{equation}
Both the additional spreading $\Delta \gamma $\ and the resonance shift $%
\Delta \omega _{0}$\ appear\ in the strength function (\ref{str}) due to the
nonlinear dissipativity effect. \bigskip

\section{Numerical results and discussion}

\bigskip We have performed the numerical calculations assuming a Fermi
distribution for the equilibrium distribution function of Eq. (\ref{f.fermi}%
) and adopting the Fermi energy $\epsilon _{F}=39$ {\rm MeV} \ and the
nuclear radius $R_{0}=r_{0}A^{1/3}$ with $r_{0}=1.12$ {\rm fm}. The higher
order relaxation parameters $\tau ^{\prime }=\zeta /mP_{{\rm eq}}$\ and $%
\tau ^{\prime \prime }=\xi /(mP_{{\rm eq}})^{2}$ are related to the
collision integral and can be evaluated using Eqs. (\ref{zetaend}) and (\ref
{xiend}) from Appendix A. We point out that in the limit of a cold nucleus, $%
T\rightarrow 0$ and $f_{{\rm eq}}=\Theta (\epsilon _{F}-\epsilon )$,\ the
corrections $\tau ^{\prime }$\ and $\tau ^{\prime \prime }$\ take the
following simple form
\begin{equation}
\frac{1}{\tau ^{\prime }}=0,\quad \frac{1}{\tau ^{\prime \prime }}\approx
1.5\ \frac{m^{3}P_{{\rm eq}}^{2}w_{0}}{\ p_{F}^{6}},  \label{tau4}
\end{equation}
where $p_{F}$ is the nucleon Fermi momentum and the scattering probability $%
w_{0}=15\pi ^{2}\hbar ^{5}/m^{3}g\alpha $ is related to the in-medium cross
section $\sigma _{{\rm in}}$ of nucleon-nucleon scattering. We use $\alpha
=9.2$ {\rm MeV} from \cite{kopl1}, which corresponds to $\sigma _{{\rm in}%
}\approx \sigma _{{\rm free}}/2,$ where $\sigma _{{\rm free}}\approx 40$
{\rm mb} is the cross section for the nucleon-nucleon scattering in free
space. Note that both relaxation parameters $\tau ^{\prime }$\ and $\tau
^{\prime \prime }$ can not be directly interpreted as the corrections to the
observable relaxation time. In particular, the value of $1/\tau ^{\prime
\prime }$ does not equal to zero in the ground state of the nucleus. The
relaxation parameters $\tau ^{\prime }$\ and $\tau ^{\prime \prime }$
determine the contribution of the viscous tensors $Q_{\alpha \beta }^{(2)}$
and $Q_{\alpha \beta }^{(3)}$ (see Eq. (\ref{q23})) to the local equations
of motion and both tensors $Q_{\alpha \beta }^{(2)}$ and $Q_{\alpha \beta
}^{(3)}$ disappear in the ground state. We also point out that the
relaxation parameters $\tau ^{\prime }$\ and $\tau ^{\prime \prime }$ as
well as $\zeta $\ and $\xi $\ depend on the nuclear mean field potential $V$%
\ due to the space integrals $r_{ij}$\ and $r_{ijk}$, see Eqs. (\ref{rij})
and (\ref{rijk}). This dependence appears after the Abrikosov-Khalatnikov
transformation (\ref{trab}) in the collision integral $\delta {\rm St}[f]$.
However, due to the presence of the strongly picked functions $\partial f_{%
{\rm eq},i}/\partial \epsilon _{i}$, at $\epsilon =\epsilon _{F}$, \ in Eqs.
(\ref{rij}) and (\ref{rijk}) the final results for $\zeta $\ and $\xi $ are
not sensitive to the specific choice of the mean field potential $V$\ at $%
T\ll \epsilon _{F}$. \ In Fig. 1, we have plotted the results of
calculations of the quantities $\hbar /\tau ^{\prime }$ (solid curve 1) and $%
\hbar /\tau ^{\prime \prime }$ (solid curve 2) as functions of temperature, $%
T$, for the nucleus with $A=224$.

\begin{figure}[tbp]
\includegraphics[width=4.0 in,height=4.0 in]{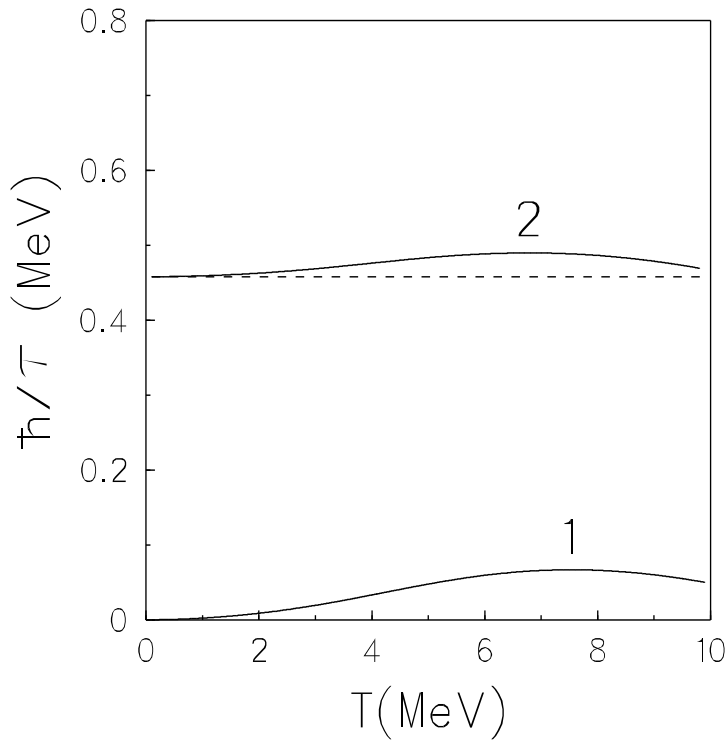}
\caption{Temperature dependence of $\hbar /\tau ^{\prime }$ (solid
curve 1) and $\hbar /\tau ^{\prime \prime }$ (solid curve 2) for
the case of the temperature-dependent Fermi distribution function
(\ref{f.fermi}). The dashed line is the calculation of $\hbar
/\tau ^{\prime \prime }$ with the sharp Thomas-Fermi distribution
function $\Theta (\epsilon _{F}-\epsilon )$.}
\end{figure}

Both quantities $\hbar /\tau ^{\prime }$ and $\hbar /\tau ^{\prime
\prime }$ show a very broad and weak maximum. The
magnitude of the maximum does not exceed the value of $0.07$ {\rm MeV} for $%
\hbar /\tau ^{\prime }$ and $0.49$ {\rm MeV} for $\hbar /\tau ^{\prime
\prime }$. The dashed line in Fig. 1 corresponds to the value of $\hbar
/\tau ^{\prime \prime }$\ from Eq. (\ref{tau4}). We can see from Fig. 1 and
Eq. (\ref{tau4}) that the simplest Thomas-Fermi distribution function $%
\Theta (\epsilon _{F}-\epsilon )$, with $\hbar /\tau ^{\prime }$
and $\hbar /\tau ^{\prime \prime }$ from Eq. (\ref{tau4}),
provides a good description of both quantities $\hbar /\tau
^{\prime }$ and $\hbar /\tau ^{\prime \prime }$. Fig. 2 shows the
ratio of the collisional relaxation time $\tau _{2}$ to both
relaxation parameters $\tau ^{\prime }$\ and $\tau ^{\prime \prime
}$. \ For the relaxation time $\tau _{2},$ we have used the
expression from Ref. \cite{kopl1} which takes into account the
memory effects. Namely,
\begin{equation}
\tau _{2}=\frac{4\pi ^{2}\alpha \hbar }{(\hbar \omega _{0})^{2}+4\pi
^{2}T^{2}}.  \label{tau5}
\end{equation}
As seen from Fig. 2, the value of $\tau _{2}/\tau ^{\prime }$ is relatively
small over the entire range of the temperature. The value of $\tau _{2}/\tau
^{\prime \prime }$ decreases with the temperature monotonically starting
from $\tau _{2}/\tau ^{\prime \prime }=$ $1.16$ at zero temperature.

\begin{figure}[tbp]
\includegraphics[width=4.0 in,height=4.0 in]{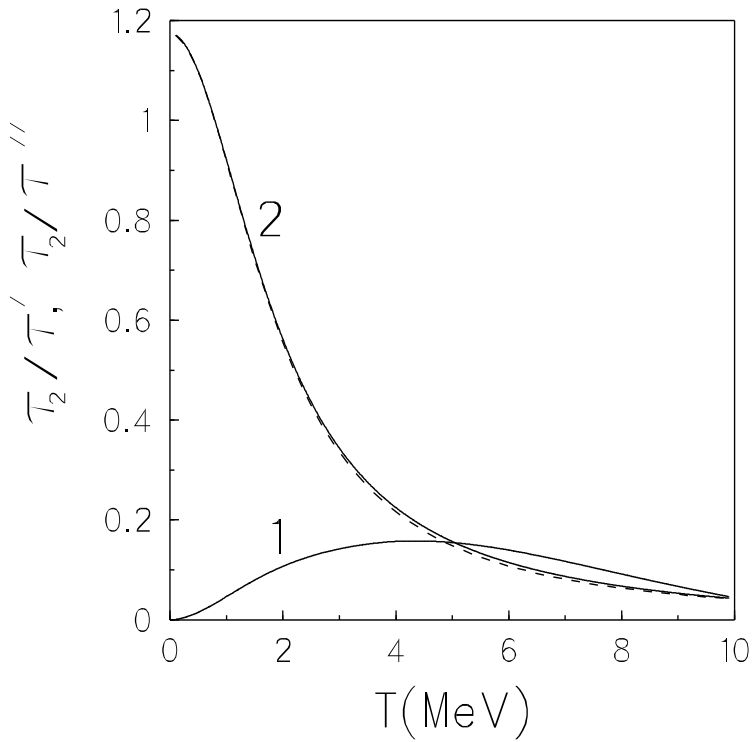}
\caption{Same as in Fig. 1 but for the ratio $\tau _{2}/\tau
^{\prime }$ multiplied by the factor $10$ (solid curve 1) and the
ratio $\tau _{2}/\tau ^{\prime \prime }$ (dashed and solid curves
2).}
\end{figure}

Let us now carry out a numerical study of the additional
contribution to the friction coefficient, $\Delta \gamma $, caused
by the nonlinear dissipativity, see Eqs. (\ref{chi4}) and
(\ref{dgg0}) and the corresponding contribution to the width
$\Gamma $ of the isoscalar giant quadrupole resonance (GQR). To
apply Eqs. (\ref{s}) and (\ref{dgg0}), we have to derive the
spectral correlation function $\left( \delta
\dot{\beta}^{2}\right) _{\omega }$. Using the correlation
properties of the random force \cite {LaLi.b.v5.76}
\begin{equation}
(\widetilde{y}^{2})_{\omega }=\frac{2\gamma _{0}T}{B}{\rm ,}  \label{y2}
\end{equation}
we obtain according to the fluctuation-dissipation theorem the following
result \cite{Kl.b.86}
\begin{equation}
\left( \delta \dot{\beta}^{2}\right) _{\omega }=\frac{2{\cal D}\ \omega ^{2}%
}{(\omega _{0}^{2}-\omega ^{2})^{2}+\gamma _{0}^{2}\omega ^{2}},
\label{xbeta}
\end{equation}
where ${\cal D}$\ is the diffusion coefficient
\begin{equation}
{\cal D}=\frac{\gamma _{0}T}{B}.  \label{fd.th}
\end{equation}
To evaluate the relative contribution to the collisional width $\Gamma $\
from $\Delta \gamma ,$\ we will start from the usual case with $\Delta
\gamma =0.$\ In this case, the width $\Gamma $\ of the GQR can be obtained
from the solution in the form $\omega =%
\mathop{\rm Re}%
\omega +i\Gamma /2\hbar $ to the following secular equation, see Eq. (\ref
{str}),
\begin{equation}
(\omega _{0}^{2}-\omega ^{2})^{2}+\gamma _{0}^{2}\omega ^{2}=0.  \label{sec1}
\end{equation}
For the numerical solution of Eq. (\ref{sec1}), we have used in Eq. (\ref
{gam0}) \ the liquid drop stiffness coefficient $C_{LDM}$ in the form \cite
{BoMo.b.v2.77}
\begin{equation}
C_{LDM}={\frac{1}{4\pi }}(L-1)(L+2)b_{S}A^{2/3}-{\frac{5}{2\pi }}{\frac{L-1}{%
2L+1}}b_{C}{\frac{Z^{2}}{A^{1/3}}},
\end{equation}
where $b_{S}=17.2$ {\rm MeV} and $b_{C}=0.7$ MeV are respectively, the
surface and Coulomb energy coefficients appearing in the nuclear mass
formula.

\begin{figure}[tbp]
\includegraphics[width=4.0 in,height=4.0 in]{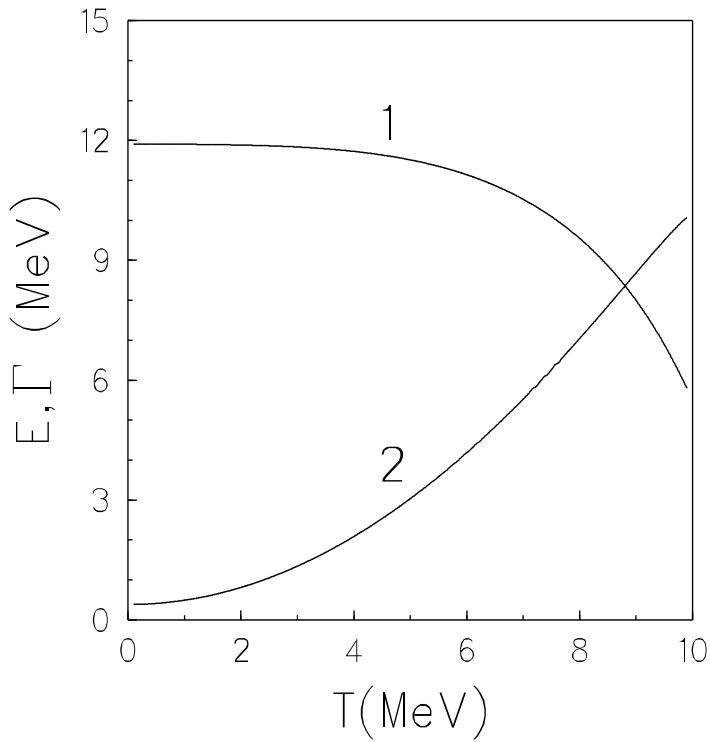}
\caption{Energy $E=\hbar
\mathop{\rm Re}%
\omega $ of the isoscalar giant quadrupole resonance (GQR) and the
corresponding collisional width $\Gamma $ for the nucleus with
$A=224$ as obtained from Eq. (\ref{sec1}).}
\end{figure}

Fig. 3 shows the results of the\ numerical solution of Eq. (\ref
{sec1}) for the nucleus with $A=224$ and $Z=A(1-6\cdot 10^{-3}A^{2/3})/2=87$%
, which corresponds to the valley of beta-stability \cite{BoMo.b.v2.77}. The
energy of the collective excitation $E=\hbar
\mathop{\rm Re}%
\omega $ decreases with temperature and approaches the hydrodynamic (liquid
drop model) limit $E_{LDM}=\hbar \sqrt{C_{LDM}/B}$ at high temperatures. In
Fig. 4 we have plotted the temperature dependence of the parameter $E\tau
_{2}/\hbar $ which determines the sound regime: $E\tau _{2}/\hbar \gg 1$\
for the zero sound (rare collision regime) and $E\tau _{2}/\hbar \ll 1$ for
the first sound (frequent collision regime).

\begin{figure}[tbp]
\includegraphics[width=4.0 in,height=4.0 in]{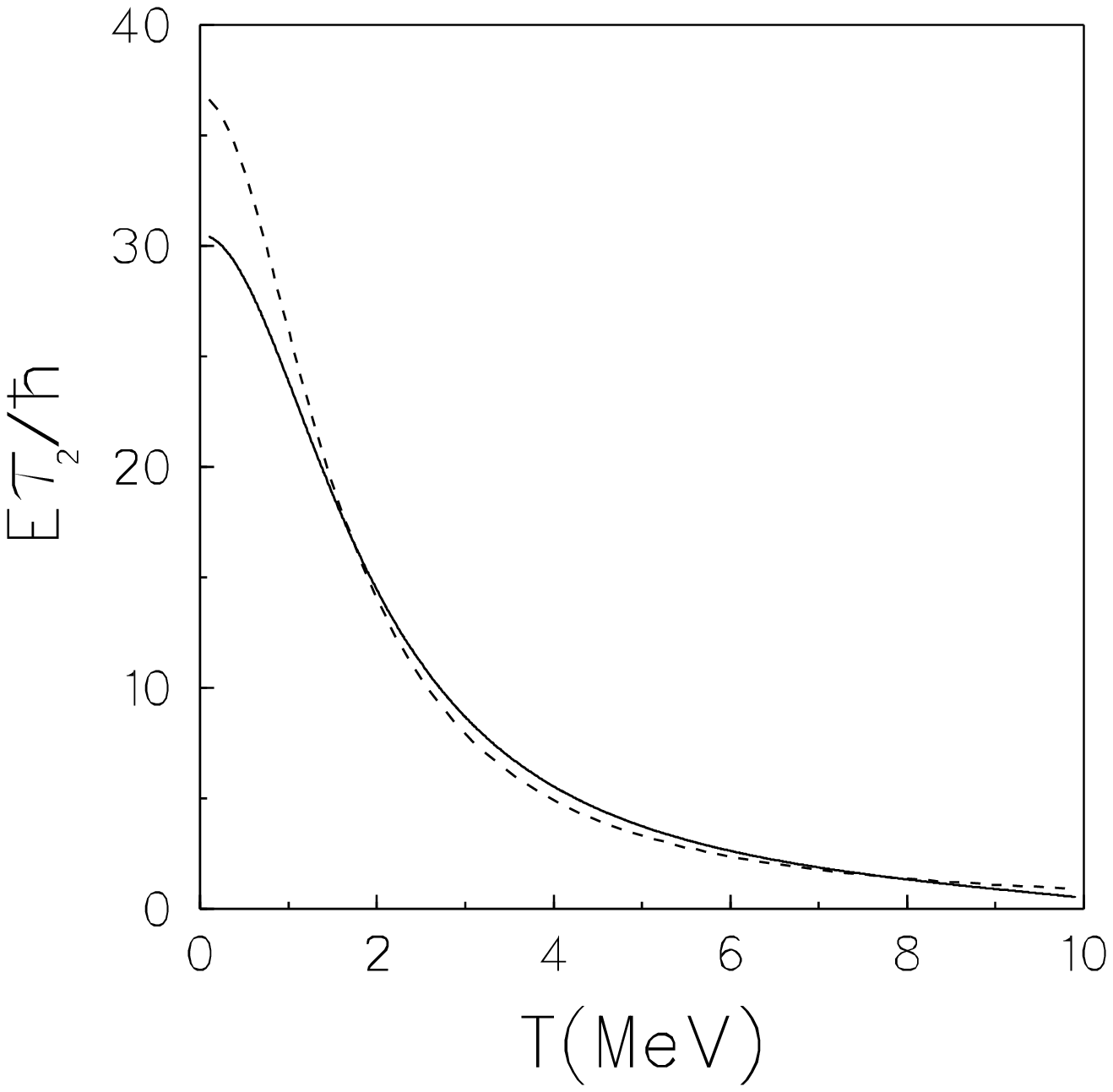}
\caption{Dependence of the dimensionless parameter $E\tau
_{2}/\hbar $ on the temperature $T$ for the GQR in the nucleus
with $A=224$ with $\tau _{2}$\ from Eq. (\ref{tau5}). The solid
curve was obtained using $E=\hbar
\mathop{\rm Re}%
\omega $ from Eq. (\ref{sec1}). The dashed curve was obtained with $E=E_{R},$%
\ where $E_{R}=60\cdot A^{-1/3}$ {\rm MeV }is the experimental
value of the GQR energy.}
\end{figure}

The solid curve in Fig. 4 corresponds to the calculation with the
temperature dependence of $E$ given by Fig. 3. For the dashed
line, the phenomenological parametrization of the
GQR energy $E=E_{R}=60\cdot A^{-1/3}$ {\rm MeV }was used. Using Eqs. (\ref{s}%
), (\ref{dgg0}), (\ref{xbeta}) and (\ref{fd.th}), one can evaluate the
contribution $\Delta \gamma $\ to the friction coefficient due to the
nonlinear dissipativity. In Fig. 5, the value of $\Delta \gamma /\gamma _{0}$%
\ is shown as a function of temperature. The ratio $\Delta \gamma /\gamma
_{0}$ equals to zero at $T=0$\ because $\Delta \gamma $\ appears due to the
thermodynamical fluctuations of the collective variable $\beta $. The ratio $%
\Delta \gamma /\gamma _{0}$ increases with temperature and reaches a maximum
value, which does not exceed $\approx 0.2$.

\begin{figure}[tbp]
\includegraphics[width=4.0 in,height=4.0 in]{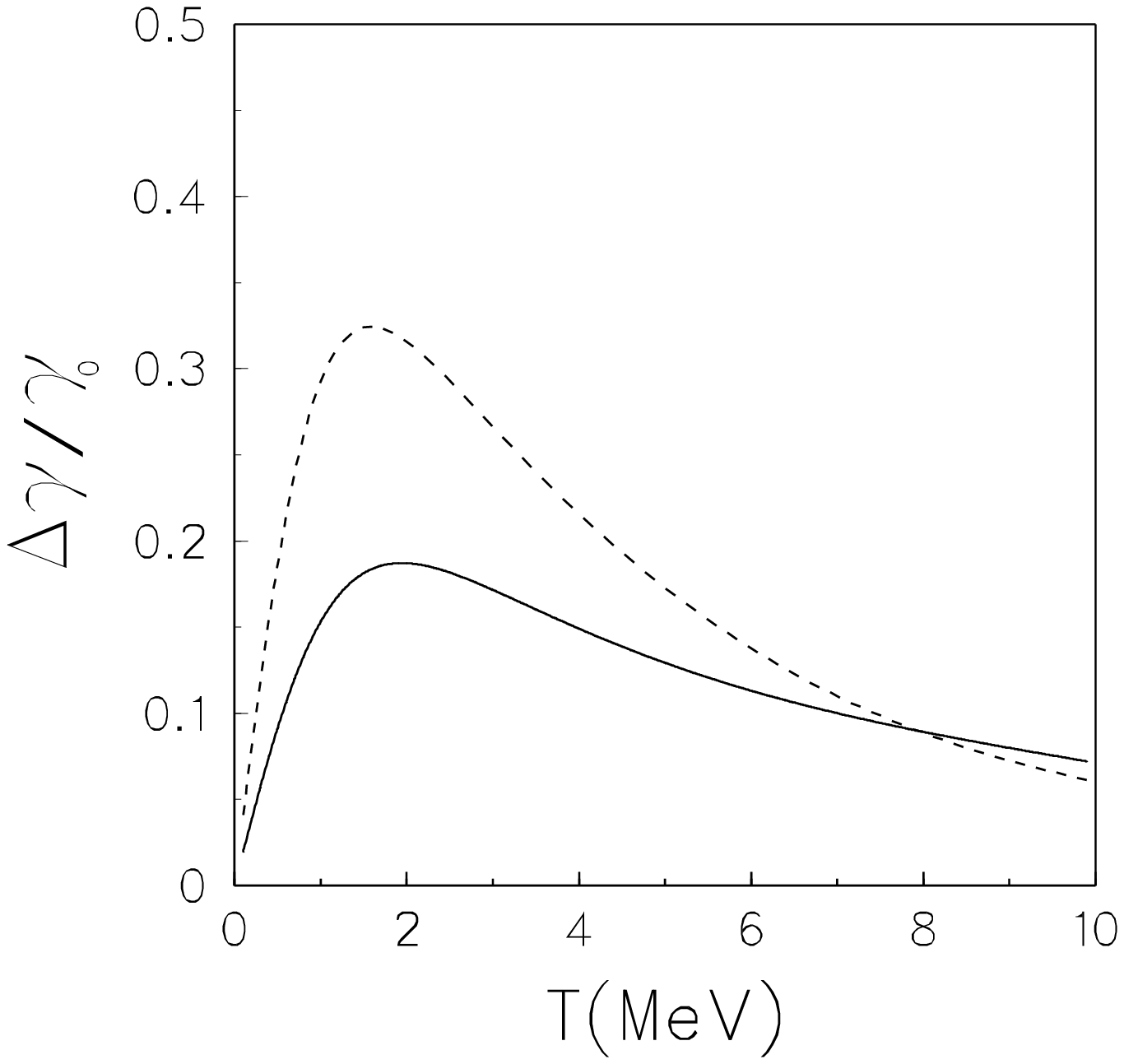}
\caption{Temperature dependence of the ratio $%
\Delta \gamma /\gamma _{0}$ for the nucleus with $A=224$ for the
GQR. The solid curve was obtained using Eq. (\ref{dgg0}) with
$\omega $ from Eq. (\ref
{sec1}) (see also Fig. 3); the dashed line was obtained using Eq. (\ref{dgg0}%
) with $\gamma _{0}$\ from Eq. (\ref{gam0}) and $\omega =\omega
_{R}=E_{R}/\hbar $ with the phenomenological parametrization
$E_{R}=60\cdot A^{-1/3}$ {\rm MeV}.}
\end{figure}

In the high-temperature region the ratio $\Delta \gamma /\gamma
_{0}$ decreases because the temperature dependence of $\gamma
_{0}\sim T^{2}$ is stronger than that of $\Delta \gamma \sim T$.
For comparison, we have also performed the calculation of the
ratio $\Delta \gamma /\gamma _{0}$ using the phenomenological
parametrization for the GQR energy $E_{R}=60\cdot A^{-1/3}$ {\rm
MeV} (see dashed line in Fig. 5). In this case, the variation of
$\Delta \gamma /\gamma _{0}$ with temperature is somewhat
stronger. Taking into account the nonlinear dissipativity effects,
the collisional width $\Gamma ^{\prime }$
of the GQR is obtained from the solution, in the form $\omega =%
\mathop{\rm Re}%
\omega +i\Gamma ^{\prime }/2\hbar ,$ to the secular equation, see Eq. (\ref
{str}),
\begin{equation}
(\omega _{0}^{2}+\Delta \omega _{0}^{2}-\omega ^{2})^{2}+(\gamma _{0}+\Delta
\gamma )^{2}\omega ^{2}=0.  \label{sec2}
\end{equation}
In Fig. 6 we have plotted the temperature dependence of the widths $\Gamma $
(dashed lines) and $\Gamma ^{\prime }$ (solid lines) for two choices of the
resonance energy: $E=\hbar
\mathop{\rm Re}%
\omega $ using Eq. (\ref{sec1}) (curves 1) and $E_{R}=60\cdot A^{-1/3}$ {\rm %
MeV} (curves 2).

\begin{figure}[tbp]
\includegraphics[width=4.0 in,height=4.0 in]{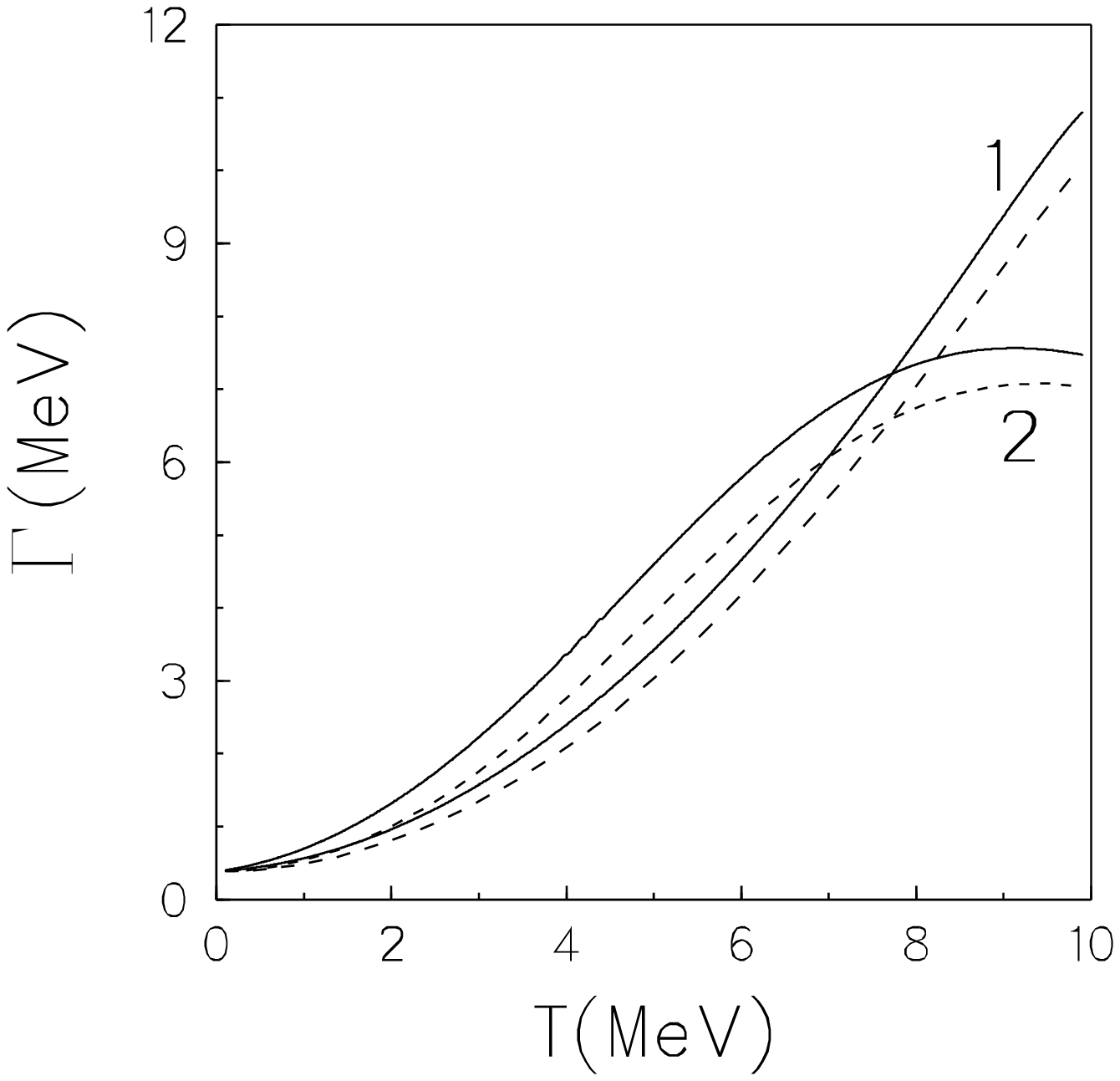}
\caption{Collisional width $\Gamma $ as a function of temperature
$T$ for the nucleus with $A=224$ for the GQR. The solid lines are
for $\Gamma =2\hbar
\mathop{\rm Im}%
\omega $ from Eq. (\ref{sec1}) and the dashed lines are for
$\Gamma ^{\prime }$\ from Eq. (\ref{sec2}). The curves 1 were
obtained using the temperature
dependent resonance frequency $\omega =\omega _{R}=%
\mathop{\rm Re}%
\omega $ from Eq. (\ref{sec1}). The curves 2 were obtained using
$\omega
=\omega _{R}=E_{R}/\hbar $ with the phenomenological parametrization $%
E_{R}=60\cdot A^{-1/3}$ {\rm MeV} (see also Fig. 5).}
\end{figure}

We point out that an increase of the width is more apparent for
curves 1 in Fig. 6 because of the temperature dependence of $E$.
The comparison of the solid and dashed lines in Fig. 6 shows that
the contribution of the nonlinear dissipative effects to the width
$\Gamma $\ does not\ exceed $\sim 20\%$.

\begin{figure}[tbp]
\includegraphics[width=4.0 in,height=4.0 in]{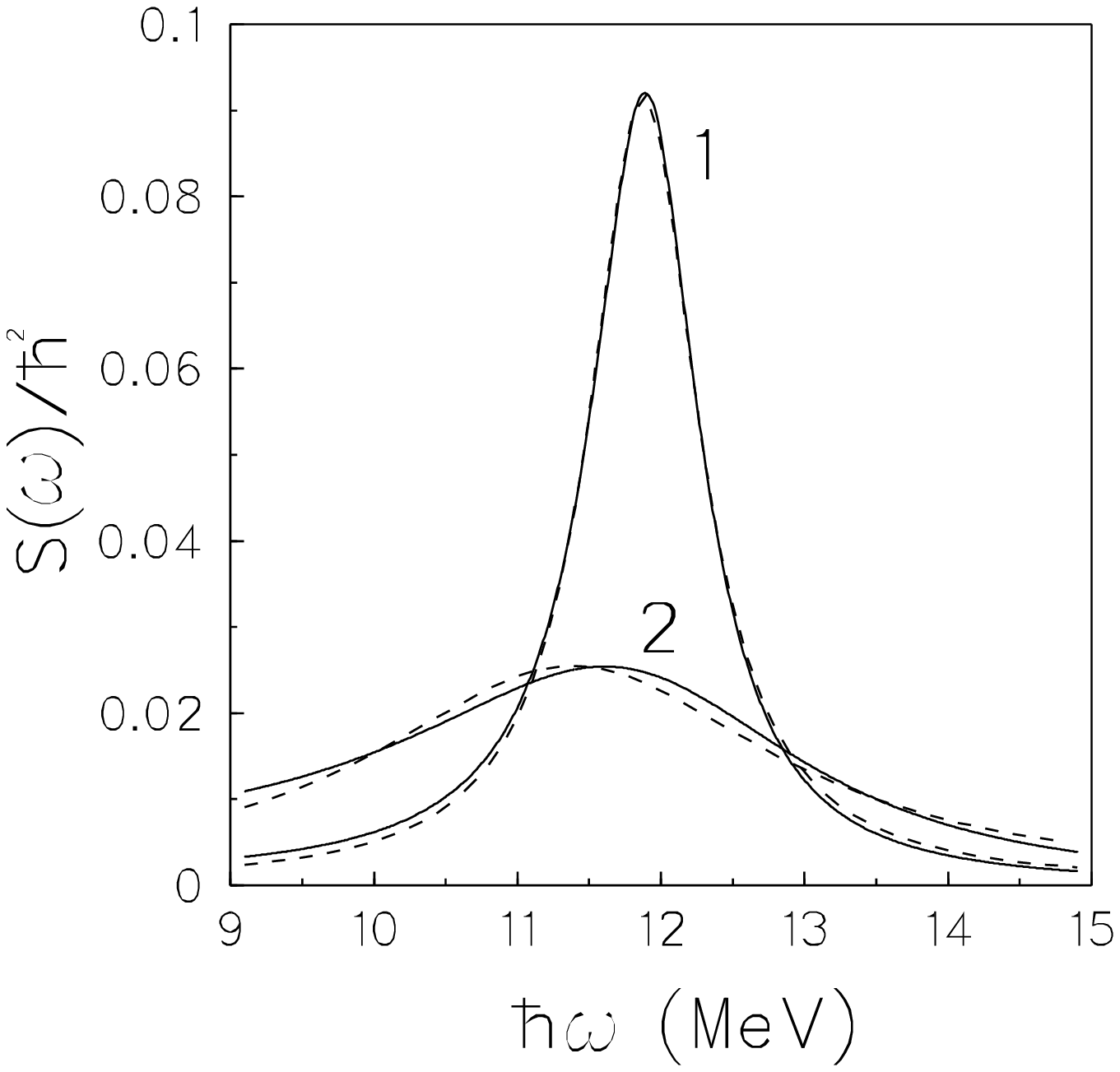}
\caption{The strength function $S(\omega )$ in $\hbar ^{2}$-units
for two temperatures: $T=1.9$ MeV (curves 1) and $T=5$ MeV (curves
2). The solid curves 1 and 2 were obtained from Eqs. (\ref{dgg0}),
\ (\ref{dww0}) and (\ref
{str}). The dashed curves 1 and 2 were obtained using Eq. (\ref{str}) with $%
\gamma _{0}+\Delta \gamma =\Gamma /\hbar $ and $\omega
_{0}^{2}+\Delta \omega _{0}^{2}=(E/\hbar )^{2}$, where $E$ and
$\Gamma $ are obtained from the solution, in the form $\omega
=E/\hbar +i\Gamma /2\hbar ,$ to the secular equation
(\ref{sec2}).}
\end{figure}

In Figs. 7 and 8 we have plotted the strength function $S(\omega
)$\ of Eq. (\ref{str}). The comparison between the solid and the
dashed lines in Fig. 7 shows the accuracy of the derivation of the
value of $\Delta \gamma $\ directly from the strength function
$S(\omega )$\ of Eq. (\ref{str}) and through the solution of the
secular equation (\ref {sec2}). The comparison of the solid lines
with the dashed lines in\ Fig. 8 demonstrates the effect of the
nonlinear dissipativity on the strength function. \bigskip

\begin{figure}[tbp]
\includegraphics[width=4.0 in,height=4.0 in]{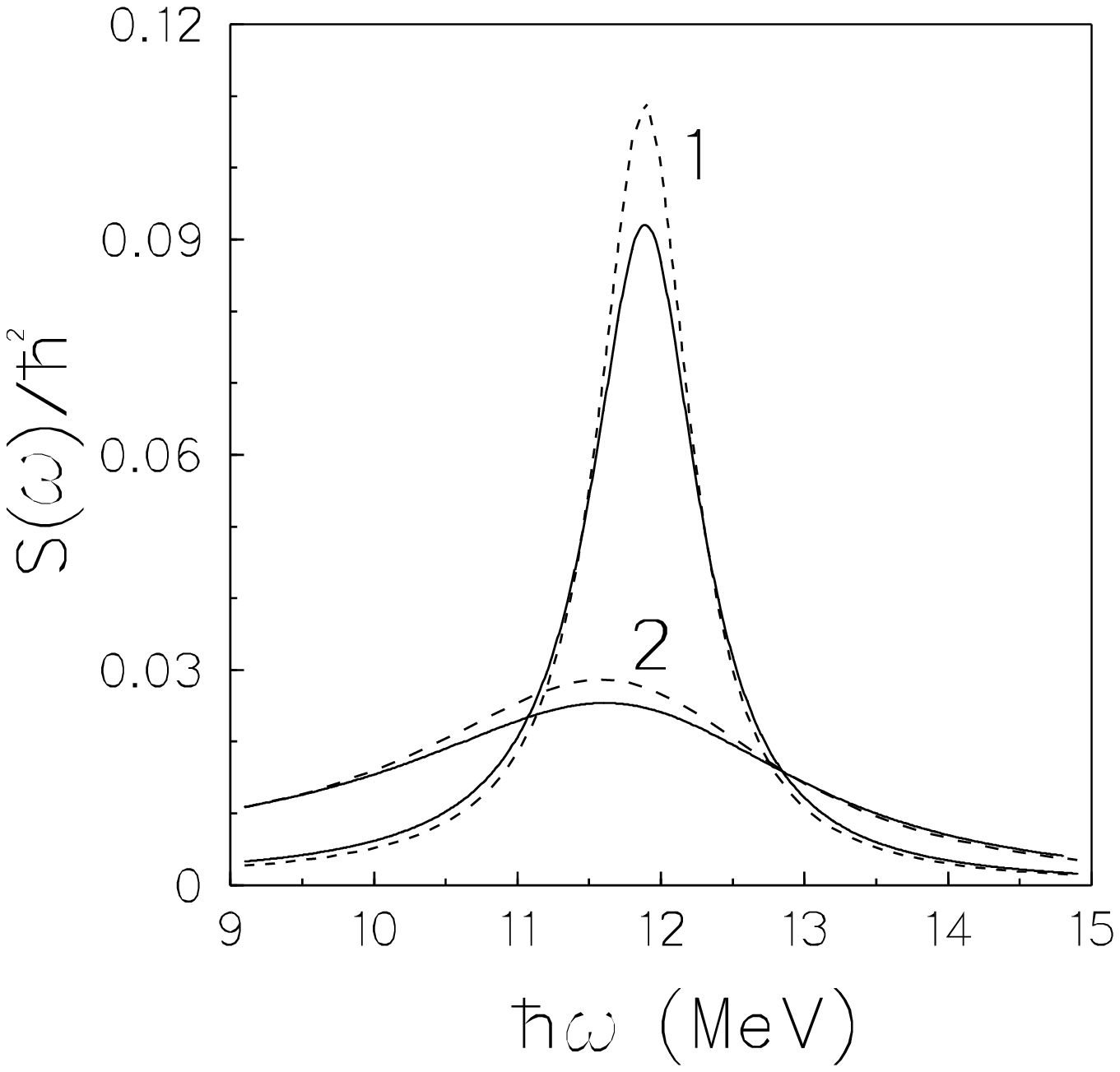}
\caption{The strength function $S(\omega )$ in $\hbar ^{2}$-units
for two temperatures: $T=1.9$ MeV (curves 1) and $T=5$ MeV (curves
2). The solid curves 1 and 2 are the same as in Fig. 7. The dashed
curves 1 and 2 were obtained from Eqs. (\ref {str}), but with
$\Delta \gamma =0$.}
\end{figure}

\section{Summary and Conclusions}

\bigskip Starting from the collisional kinetic equation with a random force
and using the ${\bf p}$-moments techniques, we have derived the equations of
motion of the viscous fluid dynamic for the local values of particle
density, velocity field and pressure tensor. The obtained equations are
closed due to the restriction imposed on the multipolarity $l$ of the Fermi
surface distortion, up to $l=2$. The important features of these equations
of motion are due to the non-Markovian form of the pressure tensor $%
P_{\alpha \beta }$. In contrast to the commonly used $\tau $-approximation,
we take into account the higher orders of the variation of the collision
integral with respect to the variation of the phase-space distribution
function. Using the Abrikosov-Khalatnikov transformation we have then
obtained the collision integral in the form of the extended $\tau $%
-approximation. Assuming a separable form for the displacement field, we
have introduced the macroscopic collective variable $\beta (t)$\ and reduced
the problem to a macroscopic equation of motion for $\beta (t)$. Note that
we do not assume the velocity$\ \dot{\beta}(t)$ to be small. The final
macroscopic equation of motion (\ref{eqbeta}) includes both the memory
effects and the nonlinear dissipativity terms $\sim \dot{\beta}^{3}$. We
have separated the description of the collective motion into two parts. The
first (slow) one is related to the driving force ${\cal F}_{{\rm ext}}(t)$
and it is associated with a slow motion having the velocity $\dot{\widetilde{%
\beta }}$. The second (fast) one is due to the random force $y(t)$\ with the
velocity $\delta \dot{\beta}\gg \dot{\widetilde{\beta }}$. Using the
correlation properties of the random force, we have performed the averaging
of the macroscopic equation of motion over the fast fluctuations $\sim
\delta \dot{\beta },$ reducing the nonlinear dissipativity terms to the form
$\sim \dot{\widetilde{\beta }}(t)\ \left\langle \delta \dot{\beta}(t^{\prime
})\delta \dot{\beta}(t^{\prime \prime })\right\rangle ,$ which is linear
with respect to the slow collective motion $\sim \dot{\widetilde{\beta }}$.
Finally, assuming a periodic driving force ${\cal F}_{{\rm ext}}(t)\sim \exp
(i\omega t)$, we have derived the macroscopic strength function $S(\omega )$%
. As seen from Eq. (\ref{str}), the nonlinear dissipativity effect leads to
the additional spreading $\Delta \gamma $\ and the resonance shift $\Delta
\omega _{0}$\ in the strength function\ $S(\omega )$. The contribution $%
\Delta \gamma $\ appears due to the thermodynamical fluctuations of the
collective variable $\beta $. In contrast to the Fermi-liquid friction
parameter $\gamma _{0}$ with $\gamma _{0}\sim T^{2}$ (at $T\ll \epsilon _{F}$%
), the spreading $\Delta \gamma $ is a linear function of the temperature $T$%
. This fact provides a non-monotonic behavior of the ratio $\Delta \gamma
/\gamma _{0}$, see Fig. 5. As seen from Fig. 5, the nonlinear dissipativity
effects are enhanced at the moderate temperatures $T\approx 2$ {\rm MeV} and
do not exceed $\approx $ 20\%. The nonlinear dissipativity effect increases
the collisional width of the GMR. Usually the total collisional width of the
isoscalar GQR in cold nuclei does not exceed 30-40\% of the experimental
value and the main contribution to the width is due to the Landau damping.
One can expect that the nonlinear dissipativity effect on the collisional
width can lead to a deviation of the temperature dependent width $\Gamma (T)$%
\ from the usual Fermi liquid prediction $\Gamma (T)\sim T^{2}$.
Unfortunately, at present time, experimental data on the temperature
behavior of $\Gamma (T)$\ of the isoscalar GQR are not available. In this
respect, it is more instructive to study the isovector giant dipole
resonance where the temperature dependence of $\Gamma (T)$\ was studied for
some heavy nuclei \cite{Ram1,Ram2}. However our final results for the
viscous tensor $Q_{\alpha \beta }$\ and the relaxation parameters $\zeta $
and $\xi $ can not be applied directly to the isovector mode because the
dipole distortion of the Fermi surface must be taken into account in the
collision integral (\ref{int1}), in contrast to our case of the isoscalar
GMR, see Sect. II. The generalization of our approach to the case isovector
modes is now in progress.

\section{Acknowledgments}

This work was supported in part by the US Department of Energy under grant
\# DOE-FG03-93ER40773. One of us (V.M.K.) thanks the Cyclotron Institute at
Texas A\&M University for the kind hospitality. \bigskip \newpage
\setcounter{equation}{0} \renewcommand{\theequation}{A\arabic{equation}}
\appendix

\begin{center}
{\bf APPENDIX A}
\end{center}

As a basic expression for the collision integral $\delta {\rm St}[f]$ we use
Eq. (\ref{int1}). The second and third variations of Eq. (\ref{int1}) with
respect to $\delta f$ take the following form
\begin{equation}
\delta {\rm St}_{2}=\int {\frac{g^{2}d\vec{p}_{2}d\vec{p}_{3}d\vec{p}_{4}}{%
(2\pi \hbar )^{6}}}w(\{\vec{p}_{j}\})\overline{\sum }\left. {\frac{\delta
^{2}Q}{\delta f(i)\delta f(j)}}\right| _{{\rm eq}}\delta f(i)\delta f(j)\
\delta (\Delta \epsilon )\delta (\Delta \vec{p}),  \label{d2st}
\end{equation}
\begin{equation}
\delta {\rm St}_{3}=\int {\frac{g^{2}d\vec{p}_{2}d\vec{p}_{3}d\vec{p}_{4}}{%
(2\pi \hbar )^{6}}}w(\{\vec{p}_{j}\})\overline{\sum }\left. {\frac{\delta
^{3}Q}{\delta f(i)\delta f(j)\delta f(k)}}\right| _{{\rm eq}}\delta
f(i)\delta f(j)\delta f(k)\ \delta (\Delta \epsilon )\delta (\Delta \vec{p}),
\label{d3st}
\end{equation}
where $\delta f(i)\equiv \delta f(\vec{p}_{i})$ and the symbol $\overline{%
\sum }$ means a summation over indices $i,j,k=1\div 4$ with $i\neq j$, $%
j\neq k$, $k\neq i$. We will follow the fluid dynamic approach and represent
the variation of the distribution function $\delta f$\ in the following
form:
\begin{equation}
\delta f(i)=-{\frac{\partial f_{{\rm eq},i}}{\partial \epsilon _{i}}}\nu
(i),\qquad \nu (i)=\sum_{l,m_{l}}^{l=2}\nu _{2m_{l}}(\vec{r}%
,t)Y_{2m_{l}}(\Omega _{i}).  \label{df}
\end{equation}
We point out that the $l=0$ and $1$ components of the expansion (\ref{df})
do not contribute to the collision integral (\ref{int1}), reflecting the
conservation of particle number and momentum in a collision.\ The expansion
coefficients $\nu _{2m}(\vec{r},t)$ in Eq. (\ref{df}) are related to the
pressure tensor $P_{\alpha \beta }^{\prime }$ of Eq. (\ref{dpi}).\ Using
Eqs. (\ref{dpi}) and (\ref{df}), we obtain
\begin{equation}
mP_{\alpha \beta }^{\prime }=-\frac{gI}{(2\pi \hbar )^{3}}%
\sum_{m_{l}=-2}^{2}\nu _{2m_{l}}\int d\Omega \ \hat{p}\newline
_{\alpha }\ \hat{p}\newline
_{\beta }Y_{2m_{l}}(\Omega ),  \label{pi5}
\end{equation}
where \
\begin{equation}
I=\int_{0}^{\infty }\ dpp^{4}\frac{\partial f_{{\rm eq}}}{\partial \epsilon }%
,  \label{i1}
\end{equation}
and $\hat{\vec{p}}=\overrightarrow{p}/p$ is the unit vector. In particular,
performing the angle integration in Eq. (\ref{pi5}), we obtain
\begin{equation}
\nu _{20}=\frac{3}{4}\sqrt{\frac{5}{\pi }}\frac{(2\pi \hbar )^{3}m}{gI}%
P_{0}^{\prime },  \label{nu2}
\end{equation}
where $P_{0}^{\prime }$\ is given by Eq. (\ref{p0}). To evaluate the
collision integral $\delta {\rm St}_{2},$\ we will substitute Eq. (\ref{df})
into (\ref{d2st}) and make use of the Abrikosov-Khalatnikov transformation
in the following form \cite{AbKh.RPP.59}
\begin{equation}
\int d\vec{p}_{2}d\vec{p}_{3}d\vec{p}_{4}\left( ...\right) \delta (\Delta
\vec{p})\Rightarrow {\frac{m^{3}}{2}}\int_{V}^{\infty }d\epsilon
_{2}d\epsilon _{3}d\epsilon _{4}\int {\frac{d\Omega d\phi _{2}}{\cos (\theta
/2)}}\left( ...\right) ,  \label{trab}
\end{equation}
where $d\Omega =\sin \theta d\theta d\phi ,$ $\theta $ is the angle between $%
\vec{p}_{1}$ and $\vec{p}_{2}$, $\phi $ is the angle between the planes
formed by ($\vec{p}_{1},\vec{p}_{2}$) and ($\vec{p}_{3},\vec{p}_{4}$),\ \
and $\phi _{2}$ is the azimuthal angle of the momentum $\vec{p}_{2}$ in the
co-ordinate system with $z$-axes along $\vec{p}_{1}$.\ We point out that the
angle $\phi $ varies only from $0$ to $\pi $ because the particles are
indistinguishable. Using\ the transformation\ (\ref{trab}) and the relation
(see Appendix B)
\begin{equation}
\int_{0}^{2\pi }{\frac{d\phi _{2}}{2\pi }}Y_{nm}(\Omega _{i})Y_{n^{\prime
}m^{\prime }}(\Omega _{j})=Y_{nm}(\Omega _{1})Y_{n^{\prime }m^{\prime
}}(\Omega _{1})P_{n}(\cos \theta _{i})P_{n^{\prime }}(\cos \theta _{j}),
\label{iyy1}
\end{equation}
we obtain
\begin{equation}
\delta {\rm St}_{2}=g^{2}{\frac{(2\pi )^{2}m^{3}}{(2\pi \hbar )^{6}}}\ [\nu
(1)]^{2}\ \overline{\sum }\langle wP_{2}(\cos \theta _{i})P_{2}(\cos \theta
_{j})\rangle r_{ij}.  \label{d2st.2}
\end{equation}
Here, $r_{ij}$ is given by
\begin{equation}
r_{ij}=\int_{V}^{\infty }d\epsilon _{2}d\epsilon _{3}d\epsilon _{4}\left. {%
\frac{\delta ^{2}Q}{\delta f(i)\delta f(j)}}\right| _{{\rm eq}}{\frac{%
\partial f_{{\rm eq},i}}{\partial \epsilon _{i}}}{\frac{\partial f_{{\rm eq}%
,j}}{\partial \epsilon _{j}}}\ \delta (\Delta \epsilon ),  \label{rij}
\end{equation}
and the symbol $\langle ...\rangle $ denotes the following average
\[
\langle w(\theta ,\phi )P_{2}(\cos \theta _{i})P_{2}(\cos \theta
_{j})\rangle =2\int_{0}^{\pi }d\theta \sin (\theta /2)\int_{0}^{\pi }\frac{%
d\phi }{2\pi }w(\theta ,\phi )P_{2}(\cos \theta _{i})P_{2}(\cos \theta _{j}),
\]
where $\cos \theta _{j}\equiv (\hat{\vec{p}}_{j}\cdot \ \hat{\vec{p}}_{1})$,
i.e. $\theta _{2}=\theta ,$ and
\begin{equation}
\cos \theta _{3}=\cos ^{2}(\theta /2)+\sin ^{2}(\theta /2)\cos \phi ,\hspace{%
0cm}\cos \theta _{4}=\cos ^{2}(\theta /2)-\sin ^{2}(\theta /2)\cos \phi ,
\label{theta}
\end{equation}
and $P_{l}(\cos \theta )$ is a Legendre polynomial. Using Eqs. (\ref{qab}), (%
\ref{int2}), (\ref{qq1q2q3}) and (\ref{d2st.2}), we obtain
\[
Q_{\alpha \beta }^{(2)}={\frac{1}{m}}\int {\frac{gd{\vec{p}}}{(2\pi \hbar
)^{3}}}(p_{\alpha }-mu_{\alpha })(p_{\beta }-mu_{\beta })\delta {\rm St}_{2}
\]
\begin{equation}
=\frac{g^{3}}{m}\frac{(2\pi )^{2}m^{3}}{(2\pi \hbar )^{9}}\overline{\sum }%
\langle wP_{2}(\cos \theta _{i})P_{2}(\cos \theta _{j})\rangle R_{ij}\ \int
d\Omega _{1}\ \hat{p}\newline
_{1,\alpha }\ \hat{p}\newline
_{1,\beta }[\nu (1)]^{2},  \label{q2.1}
\end{equation}
where
\[
R_{ij}=\int_{0}^{\infty }dp_{1}p_{1}^{4}r_{ij}.
\]
To exclude the unknown amplitude $\nu (1)$\ from Eq. (\ref{q2.1}), we will
calculate the arbitrary partial contribution to the tensor $Q_{\alpha \beta
}^{(2)}$. Using Eq. (\ref{d2st}), we will consider the partial contribution $%
Q_{\alpha \beta ,12}^{(2)}$ to the tensor $Q_{\alpha \beta }^{(2)}$\ given
by
\[
Q_{\alpha \beta ,12}^{(2)}=\frac{g^{3}}{m(2\pi \hbar )^{9}}\int d\vec{p}%
_{1}p_{1,\alpha }p_{1,\beta }\delta f(1)\int d\vec{p}_{2}\delta f(2)\int d%
\vec{p}_{3}d\vec{p}_{4}
\]
\
\begin{equation}
\times \ w(\theta ,\phi )\left. {\frac{\delta ^{2}Q}{\delta f(1)\delta f(2)}}%
\right| _{{\rm eq}}\delta (\Delta \epsilon )\delta (\Delta \vec{p}).
\label{q6}
\end{equation}
We will assume the isotropic probability scattering $w(\theta ,\phi )=w_{0}$%
, and apply the Abrikosov-Khalatnikov transformation in the following form
\begin{equation}
\int d\vec{p}_{3}d\vec{p}_{4}\left( ...\right) \delta (\Delta \vec{p}%
)\Rightarrow {\frac{m^{2}}{2p_{F}\cos (\theta /2)}}\int d\epsilon
_{3}d\epsilon _{4}d\phi \left( ...\right) ,  \label{trab1}
\end{equation}
where $p_{F}$\ is the Fermi momentum. Using Eqs. (\ref{df}) and (\ref{trab1}%
), we transform Eq. (\ref{q6}) as
\begin{equation}
Q_{\alpha \beta ,12}^{(2)}=\frac{g^{3}}{m}\frac{2\pi m^{3}w_{0}}{2(2\pi
\hbar )^{9}}R_{12}\sum_{m=-2}^{2}\nu _{2m}\int d\Omega _{2}\frac{%
Y_{2m}(\Omega _{2})}{\cos (\theta /2)}\int d\Omega _{1}\ \hat{p}\newline
_{1,\alpha }\ \hat{p}\newline
_{1,\beta }\ \nu (1).  \label{q12}
\end{equation}
Integrating over $d\Omega _{2}$\ in Eq. (\ref{q12}) and using Eq. (\ref{nu2}%
), we obtain
\begin{equation}
Q_{\alpha \beta ,12}^{(2)}=g^{2}\frac{3\pi m^{3}w_{0}}{(2\pi \hbar )^{6}}%
\frac{R_{12}}{I}P_{0}^{\prime }\int d\Omega _{1}\ \hat{p}\newline
_{1,\alpha }\ \hat{p}\newline
_{1,\beta }\ \ \nu (1).  \label{q12.1}
\end{equation}
Comparing Eq. (\ref{q12.1}) with the partial $i=1,$ $j=2$ term of Eq. (\ref
{q2.1}), we find the relation
\begin{equation}
\int d\Omega _{1}\ \hat{p}\newline
_{1,\alpha }\ \hat{p}\newline
_{1,\beta }\ [\nu (1)]^{2}=\frac{3}{4\pi g\langle P_{2}(\cos \theta )\rangle
}\frac{(2\pi \hbar )^{3}}{I}mP_{0}^{\prime }\int d\Omega _{1}\ \hat{p}%
_{1,\alpha }\ \hat{p}\newline
_{1,\beta }\ \nu (1).  \label{d2f-df}
\end{equation}
Finally, from Eqs. (\ref{q2.1}), (\ref{pi5}) and (\ref{d2f-df}) we obtain
\begin{equation}
Q_{\alpha \beta }^{(2)}=\frac{mP_{0}^{\prime }}{\zeta }P_{\alpha \beta
}^{\prime },  \label{q7y}
\end{equation}
where
\begin{equation}
\frac{1}{\zeta }=\frac{3\pi gm^{3}w_{0}}{(2\pi \hbar )^{3}}\ \frac{\overline{%
\sum }\langle P_{2}(\cos \theta _{i})P_{2}(\cos \theta _{j})\rangle R_{ij}}{%
\langle P_{2}(\cos \theta )\rangle I^{2}}.  \label{zetaend}
\end{equation}
Let us go now to the third order variation of the collision integral $\delta
{\rm St}_{3}$\ of Eq. (\ref{d3st}). Applying Eqs. (\ref{df}) and the
transformation (\ref{trab}) to Eq. (\ref{d3st}) and using the relation (see
Appendix B)
\[
\int {\frac{d\phi _{2}}{2\pi }}Y_{nm}(\Omega _{i})Y_{n^{\prime }m^{\prime
}}(\Omega _{j})Y_{n^{\prime \prime }m^{\prime \prime }}(\Omega _{k})
\]
\[
=Y_{nm}(\Omega _{1})Y_{n^{\prime }m^{\prime }}(\Omega _{1})Y_{n^{\prime
\prime }m^{\prime \prime }}(\Omega _{1})P_{n}(\cos \theta _{i})P_{n^{\prime
}}(\cos \theta _{j})P_{n^{\prime \prime }}(\cos \theta _{k}),
\]
we will reduce the collision integral $\delta {\rm St}_{3}$\ to the
following form
\begin{equation}
\delta {\rm St}_{3}=-g^{2}\frac{(2\pi )^{2}m^{3}}{(2\pi \hbar )^{6}}\ [\nu
(1)]^{3}\ \overline{\sum }\langle wP_{2}(\cos \theta _{i})P_{2}(\cos \theta
_{j})P_{2}(\cos \theta _{k})\rangle r_{ijk},  \label{d3st.2}
\end{equation}
where
\begin{equation}
r_{ijk}=\int_{V}^{\infty }d\epsilon _{2}d\epsilon _{3}d\epsilon _{4}\left. {%
\frac{\delta ^{3}Q}{\delta f(i)\delta f(j)\delta f(k)}}\right| _{eq}{\frac{%
\partial f_{{\rm eq},i}}{\partial \epsilon _{i}}}{\frac{\partial f_{{\rm eq}%
,j}}{\partial \epsilon _{j}}}{\frac{\partial f_{{\rm eq},k}}{\partial
\epsilon _{k}}}\ \delta (\Delta \epsilon ).  \label{rijk}
\end{equation}
Using Eqs. (\ref{qab}), (\ref{int2}), (\ref{qq1q2q3}) and (\ref{d3st.2}), we
obtain
\[
Q_{\alpha \beta }^{(3)}={\frac{1}{m}}\int {\frac{gd{\vec{p}}}{(2\pi \hbar
)^{3}}}(p_{\alpha }-mu_{\alpha })(p_{\beta }-mu_{\beta })\delta {\rm St}_{3}
\]
\begin{equation}
=-\frac{g^{3}}{m}\frac{(2\pi )^{2}m^{3}}{(2\pi \hbar )^{9}}\overline{\sum }%
\langle wP_{2}(\cos \theta _{i})P_{2}(\cos \theta _{j})P_{2}(\cos \theta
_{k})\rangle R_{ijk}\ \int d\Omega _{1}\ \hat{p}_{1,\alpha }\ \hat{p}%
_{1,\beta }\ \ [\nu (1)]^{3},  \label{q3.2}
\end{equation}
where
\begin{equation}
R_{ijk}=\int_{0}^{\infty }dpp^{4}r_{ijk}.
\end{equation}
Similar to the previous evaluation of the tensor $Q_{\alpha \beta }^{(2)}$,
we will consider the partial term of Eq. (\ref{d3st.2}) with $i=1,\ j=2,\ k=3
$\ and the corresponding partial tensor $Q_{\alpha \beta ,123}^{(3)}$\ which
is given by
\[
Q_{\alpha \beta ,123}^{(3)}=-\frac{g^{3}}{m(2\pi \hbar )^{9}}\int d\vec{p}%
_{1}\ \hat{p}_{1,\alpha }\hat{p}_{1,\beta }\ \delta f(1)
\]
\begin{equation}
\times \int d\vec{p}_{2}\delta f(2)\int d\vec{p}_{3}\delta f(3)\int d\vec{p}%
_{4}w(\theta ,\phi )\delta (\Delta \epsilon )\delta (\Delta \vec{p}),
\label{q3x}
\end{equation}
where we have used the following relation
\[
\left. {\frac{\delta ^{3}Q}{\delta f(1)\delta f(2)\delta f(3)}}\right| _{%
{\rm eq}}=-1.
\]
We will again assume the isotropic scattering probability: $w(\theta ,\phi
)=w_{0}$ and apply the transformation (\ref{trab1}) to Eq. (\ref{q3x}). The\
angle integrals over $d\Omega _{2}d\phi ,$ appearing in Eq. (\ref{q3x}), can
be transformed as
\begin{equation}
\int d\Omega _{2}d\phi \frac{Y_{2m^{\prime }}(\Omega _{2})}{\cos (\theta /2)}%
Y_{2m^{\prime \prime }}(\Omega _{3})=Y_{2m^{\prime \prime }}(\Omega
_{1})\int d\Omega _{2}d\phi \frac{P_{2}(\cos \theta _{3})}{\cos (\theta /2)}%
Y_{2m^{\prime }}(\Omega _{2}),  \label{rel}
\end{equation}
where we have used the relation \cite{BrSy.AP.70}
\begin{equation}
\int_{0}^{\pi }{\frac{d\phi _{2}}{2\pi }}Y_{nm}(\Omega _{i})=Y_{nm}(\Omega
_{1})P_{n}(\cos \theta _{i}).  \label{y1}
\end{equation}
The result reads
\[
Q_{\alpha \beta ,123}^{(3)}=-\frac{g^{3}}{m}\frac{m^{3}w_{0}}{2(2\pi \hbar
)^{9}}R_{123}\sum_{m=-2}^{2}\nu _{2m}
\]
\begin{equation}
\times \int d\Omega _{2}d\phi \frac{P_{2}(\cos \theta _{3})}{\cos (\theta /2)%
}Y_{2m}(\Omega _{2})\int d\Omega _{1}\ \hat{p}\newline
_{1,\alpha }\ \hat{p}\newline
_{1,\beta }\ [\nu (1)]^{2}.  \label{q3.s}
\end{equation}
Performing the integration over $d\Omega _{2}d\phi $\ in Eq. (\ref{q3.s})
and using Eqs. (\ref{d2f-df}) and (\ref{nu2}), we obtain
\begin{equation}
Q_{\alpha \beta ,123}^{(3)}=-\frac{27}{28}\frac{gm^{3}}{(2\pi \hbar )^{3}}%
\frac{w_{0}}{\langle P_{2}(\cos \theta )\rangle }\frac{R_{123}}{I^{2}}%
mP_{0}^{\prime 2}\int d\Omega _{1}\ \hat{p}\newline
_{1,\alpha }\ \hat{p}\newline
_{1,\beta }\ \nu (1).  \label{q3.3}
\end{equation}
Comparing Eq. (\ref{q3.3}) with the partial $i=1,\ j=2,\ k=3$ term of Eq. (%
\ref{q3.2}) we obtain the following relation
\[
\int d\Omega _{1}\ \hat{p}_{1,\alpha }\ \hat{p}_{1,\beta }\ \nu ^{3}(1)=%
\frac{27}{28}\frac{(2\pi \hbar )^{6}}{(2\pi )^{2}g^{2}I^{2}}\frac{%
m^{2}P_{0}^{\prime 2}}{\langle P_{2}(\cos \theta )\rangle \langle P_{2}(\cos
\theta )P_{2}(\cos \theta _{3})\rangle }
\]
\begin{equation}
\times \int d\Omega _{1}\ \hat{p}\newline
_{1,\alpha }\ \hat{p}\newline
_{1,\beta }\ \nu (1).  \label{d3f-d2f}
\end{equation}
Finally, substituting Eq. (\ref{d3f-d2f}) into Eq. (\ref{q3.2})\ and using
Eq. (\ref{pi5}), we obtain
\begin{equation}
Q_{\alpha \beta }^{(3)}=\frac{m^{2}P_{0}^{\prime 2}}{\xi }P_{\alpha \beta
}^{\prime },  \label{q7x}
\end{equation}
where
\begin{equation}
{\frac{1}{\xi }}=\frac{27m^{3}w_{0}}{28}\ \frac{\overline{\sum }\langle
P_{2}(\cos \theta _{i})P_{2}(\cos \theta _{j})P_{2}(\cos \theta _{k})\rangle
R_{ijk}}{\langle P_{2}(\cos \theta )\rangle \langle P_{2}(\cos \theta
)P_{2}(\cos \theta _{3})\rangle I^{3}}.  \label{xiend}
\end{equation}
\setcounter{equation}{0} \renewcommand{\theequation}{B\arabic{equation}}
\appendix

\begin{center}
{\bf APPENDIX B}
\end{center}

\bigskip In this Appendix, we will consider some angle integrals which
appear in the calculations of the collision integral and its variations. Let
us start from the integral
\begin{equation}
M_{23}=\int \frac{d\Omega _{2}}{4\pi }P_{n}^{m}(\cos \Theta _{2})e^{im\Phi
_{2}}P_{n^{\prime }}^{m^{\prime }}(\cos \Theta _{3})e^{im\Phi
_{3}}P_{l}(\cos \theta ),  \label{m23}
\end{equation}
where $(\Theta _{j},\Phi _{j})$ are the angle coordinates of the momentum
vectors $\vec{p}_{j}$ in the arbitrary co-ordinate frame ($j=1\div 4)$ and $%
\theta $\ is the angle between the vectors $\vec{p}_{1}$ and $\vec{p}_{2}$.
Using the addition theorem for spherical harmonics \cite{BeEr.b.65}
\begin{equation}
P_{l}(\cos \theta )=\sum_{r=-l}^{l}\frac{(l-|r|)!}{(l+|r|)!}P_{l}^{|r|}(\cos
\Theta _{1})P_{l}^{|r|}(\cos \Theta _{2})e^{ir(\Phi _{1}-\Phi _{2})},
\label{pl}
\end{equation}
we find
\[
M_{23}=\sum_{r=-l}^{l}\frac{(l-|r|)!}{(l+|r|)!}\int \frac{d\Phi _{2}}{4\pi }%
\int \sin \Theta _{2}d\Theta _{2}P_{n}^{m}(\cos \Theta _{2})P_{n^{\prime
}}^{m^{\prime }}(\cos \Theta _{3})
\]
\[
\times P_{l}^{|r|}(\cos \Theta _{2})P_{l}^{|r|}(\cos \Theta _{1})e^{im\Phi
_{2}}e^{im^{\prime }\Phi _{3}}e^{ir(\Phi _{1}-\Phi _{2})}
\]
\begin{equation}
=\frac{\delta _{nl}}{2n+1}P_{n}^{m}(\cos \Theta _{1})e^{im\Phi
_{1}}P_{n^{\prime }}^{m^{\prime }}(\cos \Theta _{1})e^{im^{\prime }\Phi
_{1}}P_{n^{\prime }}(\cos \Theta _{3}).  \label{m23-1}
\end{equation}
Here, $\theta _{3}$ is the angle between $\vec{p}_{1}$ and $\vec{p}_{3}$. On
the other hand, using the direction of $\vec{p}_{1}$ as a polar axis with $%
d\Omega _{2}=\sin \theta d\theta d\phi _{2}$ where $\phi _{2}$ is the
azimuthal coordinate of $\vec{p}_{2}$ in the new co-ordinate frame, we will
rewrite Eq. (\ref{m23}) as
\begin{equation}
M_{23}=\int \frac{d\theta }{2}\sin \theta P_{l}(\cos \theta )\left\{ \int
\frac{d\phi _{2}}{2\pi }P_{n}^{m}(\cos \Theta _{2})e^{im\Phi
_{2}}P_{n^{\prime }}^{m^{\prime }}(\cos \Theta _{3})e^{im^{\prime }\Phi
_{3}}\right\} .  \label{m23-x}
\end{equation}
Here and below the angles \ $\Theta _{i}$ and $\Phi _{i}$\ are dependent on
the angles $\theta $\ and $\phi _{2}.$\ Comparing Eqs. (\ref{m23-1}) and (%
\ref{m23-x}) and using the orthogonality condition for the Legendre
polynomial $P_{l}(\cos \theta )$\ in Eq. (\ref{m23-x}), one obtains the
following integral relation
\[
\int \frac{d\phi _{2}}{2\pi }P_{n}^{m}(\cos \Theta _{2})e^{im\Phi
_{2}}P_{n^{\prime }}^{m^{\prime }}(\cos \Theta _{3})e^{im^{\prime }\Phi _{3}}
\]
\begin{equation}
=P_{n}^{m}(\cos \Theta _{1})e^{im\Phi _{1}}P_{n^{\prime }}^{m^{\prime
}}(\cos \Theta _{1})e^{im^{\prime }\Phi _{1}}P_{n}(\cos \theta )P_{n^{\prime
}}(\cos \theta _{3}).  \label{m23-0}
\end{equation}
Starting from the integral
\[
M_{24}=\int \frac{d\Omega _{2}}{4\pi }P_{n}^{m}(\cos \Theta _{2})e^{im\Phi
_{2}}P_{n^{\prime }}^{m^{\prime }}(\cos \Theta _{4})e^{im\Phi
_{4}}P_{l}(\cos \theta ),
\]
we will also obtain an integral relation analogous to Eq. (\ref{m23-0}) but
with the replacement $3\rightarrow 4.$ Let us consider now the integral
\begin{equation}
M_{34}=\int \frac{d\Omega _{3}}{4\pi }P_{n}^{m}(\cos \Theta _{3})e^{im\Phi
_{3}}P_{n^{\prime }}^{m^{\prime }}(\cos \Theta _{4})e^{im^{\prime }\Phi
_{4}}P_{l}(\cos \theta _{3}).  \label{m34}
\end{equation}
Using the addition theorem for $P_{l}(\cos \theta _{3})$ (see Eq. (\ref{pl}%
)), we reduce Eq. (\ref{m34}) as
\[
M_{34}=\frac{\delta _{nl}}{2n+1}P_{n}^{m}(\cos \Theta _{1})e^{im\Phi
_{1}}\int \frac{d\Phi _{3}}{2\pi }P_{n^{\prime }}^{m^{\prime }}(\cos \Theta
_{4})e^{im^{\prime }\Phi _{4}}
\]
\begin{equation}
=\frac{\delta _{nl}}{2n+1}P_{n}^{m}(\cos \Theta _{1})e^{im\Phi
_{1}}P_{n^{\prime }}^{m^{\prime }}(\cos \Theta _{1})e^{im^{\prime }\Phi
_{1}}P_{n}(\cos \Theta _{4}).  \label{m34-x}
\end{equation}
Replacing in Eq. (\ref{m34}) the integration over $\Phi _{3}$ to integration
over $\Phi _{2}$\ and using the direction\ of $\vec{p}_{1}$ as a polar axis,
we will rewrite Eq. (\ref{m34}) as
\begin{equation}
M_{34}=\int \frac{d\theta _{3}}{2}\sin \theta _{3}P_{l}(\cos \theta
_{3})\left\{ \int \frac{d\phi _{2}}{2\pi }P_{n}^{m}(\cos \Theta
_{3})e^{im\Phi _{3}}P_{n^{\prime }}^{m^{\prime }}(\cos \Theta
_{4})e^{im^{\prime }\Phi _{4}}\right\} .  \label{m34-y}
\end{equation}
Comparing Eqs. (\ref{m34-x}) and (\ref{m34-y}) and using the orthogonality
conditions for the Legendre polynomials, we obtain
\[
\int \frac{d\phi _{2}}{2\pi }P_{n}^{m}(\cos \Theta _{3})e^{im\Phi
_{3}}P_{n^{\prime }}^{m^{\prime }}(\cos \Theta _{4})e^{im^{\prime }\Phi _{4}}
\]
\begin{equation}
=P_{n}^{m}(\cos \Theta _{1})e^{im\Phi _{1}}P_{n^{\prime }}^{m^{\prime
}}(\cos \Theta _{1})e^{im^{\prime }\Phi _{1}}P_{n}(\cos \theta
_{3})P_{n^{\prime }}(\cos \theta _{4}).  \label{m34-0}
\end{equation}
Using the representation of the spherical function $Y_{nm}(\Omega )$\ via
the Legendre polynomials $P_{n}^{m}(\cos \theta )$ \cite{BeEr.b.65}
\begin{equation}
Y_{nm}(\Omega )=\sqrt{\frac{2n+1}{4\pi }\frac{(n-m)!}{(n+m)!}}P_{n}^{m}(\cos
\theta )e^{im\phi },  \label{yp}
\end{equation}
and collecting Eqs. (\ref{m23-0}) and (\ref{m34-0}) we obtain the following
integral relation
\begin{equation}
\int \frac{d\phi _{2}}{2\pi }Y_{nm}(\Omega _{i})Y_{n^{\prime }m^{\prime
}}(\Omega _{j})=Y_{nm}(\Omega _{1})Y_{n^{\prime }m^{\prime }}(\Omega
_{1})P_{n}(\cos \theta _{i})P_{n^{\prime }}(\cos \theta _{j}),  \label{m2-0}
\end{equation}
where $i,j=1\div 4$. \bigskip Let us consider finally the integral
\begin{equation}
M_{234}=\int \frac{d\Omega _{2}}{4\pi }P_{n}^{m}(\cos \Theta _{2})e^{im\Phi
_{2}}P_{n^{\prime }}^{m^{\prime }}(\cos \Theta _{3})e^{im\Phi
_{3}}P_{n^{\prime \prime }}^{m^{\prime \prime }}(\cos \Theta _{4})e^{im\Phi
_{4}}P_{l}(\cos \theta ).  \label{m234}
\end{equation}
Similar to the previous consideration, we will transform Eq. (\ref{m234}) as
\[
M_{234}=\int \frac{d\theta }{2}\sin \theta P_{l}(\cos \theta )
\]
\[
\times \left\{ \int \frac{d\phi _{2}}{2\pi }P_{n}^{m}(\cos \Theta
_{2})e^{im\Phi _{2}}P_{n^{\prime }}^{m^{\prime }}(\cos \Theta
_{3})e^{im^{\prime }\Phi _{3}}P_{n^{\prime \prime }}^{m^{\prime \prime
}}(\cos \Theta _{4})e^{im^{\prime }\Phi _{4}}\right\}
\]
\[
=\frac{\delta _{nl}}{2n+1}P_{n}^{m}(\cos \Theta _{1})e^{im\Phi
_{1}}P_{n^{\prime }}^{m^{\prime }}(\cos \Theta _{1})e^{im^{\prime }\Phi _{1}}
\]
\begin{equation}
\times P_{n^{\prime \prime }}^{m^{\prime \prime }}(\cos \Theta
_{1})e^{im^{\prime \prime }\Phi _{1}}P_{n^{\prime }}(\cos \Theta
_{3})P_{n^{\prime \prime }}(\cos \Theta _{4}).
\end{equation}
Using the orthogonality conditions for the Legendre polynomials, we obtain
\[
\int \frac{d\phi _{2}}{2\pi }P_{n}^{m}(\cos \Theta _{2})e^{im\Phi
_{2}}P_{n^{\prime }}^{m^{\prime }}(\cos \Theta _{3})e^{im^{\prime }\Phi
_{3}}P_{n^{\prime \prime }}^{m^{\prime \prime }}(\cos \Theta
_{4})e^{im^{\prime \prime }\Phi _{4}}
\]
\[
=P_{n}^{m}(\cos \Theta _{1})e^{im\Phi _{1}}P_{n^{\prime }}^{m^{\prime
}}(\cos \Theta _{1})e^{im^{\prime }\Phi _{1}}P_{n^{\prime \prime
}}^{m^{\prime \prime }}(\cos \Theta _{1})e^{im^{\prime \prime }\Phi _{1}}
\]
\begin{equation}
\times P_{n}(\cos \theta )P_{n^{\prime }}(\cos \theta _{3})P_{n^{\prime
\prime }}(\cos \theta _{4}).  \label{m234-0}
\end{equation}
Finally, taking into account Eq. (\ref{yp}) we will generalize Eq. (\ref
{m234-0}) as
\[
\int \frac{d\phi _{2}}{2\pi }Y_{nm}(\Omega _{i})Y_{n^{\prime }m^{\prime
}}(\Omega _{j})Y_{n^{\prime \prime }m^{\prime \prime }}(\Omega _{k})
\]
\begin{equation}
=Y_{nm}(\Omega _{1})Y_{n^{\prime }m^{\prime }}(\Omega _{1})Y_{n^{\prime
\prime }m^{\prime \prime }}(\Omega _{1})P_{n}(\cos \theta _{i})P_{n^{\prime
}}(\cos \theta _{j})P_{n^{\prime \prime }}(\cos \theta _{k}).  \label{m3-0}
\end{equation}
\setcounter{equation}{0} \renewcommand{\theequation}{C\arabic{equation}}
\appendix

\begin{center}
{\bf APPENDIX C}
\end{center}

In this Appendix we give a proof of the macroscopic equation of motion for
the nuclear shape variable $\beta (t)$ derived by the displacement field as $%
\vec{\chi}(\vec{r},t)=\beta (t)\vec{v}(\vec{r})$, see Sect. III.
Substituting this separable form in Eq. (\ref{sys2.2}) and multiplying by $%
v_{\alpha }$, summing over $\alpha $, and integrating over $\vec{r}$ space,
we obtain the equation of motion for the collective variable $\beta (t).$\
Namely,
\[
B\ddot{\beta}+D_{0}\dot{\beta}^{2}+A_{0}\int_{-\infty }^{t}dt^{\prime }\exp
\left( \frac{t^{\prime }-t}{\tau _{2}}\right) \ \dot{\beta}(t^{\prime
})+C_{LDM}\beta
\]
\[
-D_{1}\int_{-\infty }^{t}dt^{\prime }\int_{-\infty }^{t^{\prime
}}dt_{1}^{\prime }\ \exp \left( \frac{t_{1}^{\prime }-t}{\tau _{2}}\right) \
\dot{\beta}(t^{\prime })\dot{\beta}(t_{1}^{\prime })
\]
\[
-D_{2}\int_{-\infty }^{t}dt^{\prime }\int_{-\infty }^{t^{\prime
}}dt_{1}^{\prime }\int_{-\infty }^{t^{\prime }}dt_{2}^{\prime }\exp \left(
\frac{t_{1}^{\prime }-t}{\tau _{2}}\right) \exp \left( \frac{t_{2}^{\prime
}-t^{\prime }}{\tau _{2}}\right) \ \dot{\beta}(t_{1}^{\prime })\dot{\beta}%
(t_{2}^{\prime })
\]
\[
+A_{1}\int_{-\infty }^{t}dt^{\prime }\int_{-\infty }^{t^{\prime
}}dt_{1}^{\prime }\int_{-\infty }^{t^{\prime }}dt_{2}^{\prime }\int_{-\infty
}^{t^{\prime }}dt_{3}^{\prime }\exp \left( \frac{t_{1}^{\prime }-t}{\tau _{2}%
}\right) \exp \left( \frac{t_{2}^{\prime }-t^{\prime }}{\tau _{2}}\right)
\]
\[
\times \exp \left( \frac{t_{3}^{\prime }-t^{\prime }}{\tau _{2}}\right) \
\dot{\beta}(t_{1}^{\prime })\dot{\beta}(t_{2}^{\prime })\dot{\beta}%
(t_{3}^{\prime })
\]
\[
+A_{2}\int_{-\infty }^{t}dt^{\prime }\int_{-\infty }^{t^{\prime
}}dt_{1}^{\prime }\int_{-\infty }^{t_{1}^{\prime }}dt_{2}^{\prime
}\int_{-\infty }^{t^{\prime }}dt_{3}^{\prime }\exp \left( \frac{%
t_{2}^{\prime }-t}{\tau _{2}}\right) \exp \left( \frac{t_{3}^{\prime
}-t^{\prime }}{\tau _{2}}\right) \ \dot{\beta}(t_{1}^{\prime })\dot{\beta}%
(t_{2}^{\prime })\dot{\beta}(t_{3}^{\prime })
\]
\[
+A_{3}\int_{-\infty }^{t}dt^{\prime }\int_{-\infty }^{t^{\prime
}}dt_{1}^{\prime }\int_{-\infty }^{t_{1}^{\prime }}dt_{2}^{\prime
}\int_{-\infty }^{t_{1}^{\prime }}dt_{3}^{\prime }\int_{-\infty }^{t^{\prime
}}dt_{4}^{\prime }\exp \left( \frac{t_{2}^{\prime }-t}{\tau _{2}}\right)
\exp \left( \frac{t_{3}^{\prime }-t_{1}^{\prime }}{\tau _{2}}\right)
\]
\[
\times \exp \left( \frac{t_{4}^{\prime }-t^{\prime }}{\tau _{2}}\right) \dot{%
\beta}(t_{2}^{\prime })\dot{\beta}(t_{3}^{\prime })\dot{\beta}(t_{4}^{\prime
})
\]
\[
+A_{4}\int_{-\infty }^{t}dt^{\prime }\int_{-\infty }^{t^{\prime
}}dt_{1}^{\prime }\int_{-\infty }^{t_{1}^{\prime }}dt_{2}^{\prime }\exp
\left( \frac{t_{2}^{\prime }-t}{\tau _{2}}\right) \ \dot{\beta}(t^{\prime })%
\dot{\beta}(t_{1}^{\prime })\dot{\beta}(t_{2}^{\prime })
\]
\[
+A_{5}\int_{-\infty }^{t}dt^{\prime }\int_{-\infty }^{t^{\prime
}}dt_{1}^{\prime }\int_{-\infty }^{t_{1}^{\prime }}dt_{2}^{\prime
}\int_{-\infty }^{t_{1}^{\prime }}dt_{3}^{\prime }\ \exp \left( \frac{%
t_{2}^{\prime }-t}{\tau _{2}}\right) \exp \left( \frac{t_{3}^{\prime
}-t_{1}^{\prime }}{\tau _{2}}\right) \ \dot{\beta}(t^{\prime })\dot{\beta}%
(t_{2}^{\prime })\dot{\beta}(t_{3}^{\prime })
\]
\begin{equation}
=B\ {\cal F}_{{\rm ext}}(t)+B\ \widetilde{y}(t),  \label{eqbeta}
\end{equation}
where $B{\cal F}_{{\rm ext}}$\ and $B\ \widetilde{y}(t)$ are, respectively,
the external and random forces in the collective space of the variable $%
\beta $\ (we have separated the mass coefficient $B$\ from the external and
random forces for technical convenience). The transport coefficients in Eq. (%
\ref{eqbeta}) are given by
\begin{equation}
B=m\int d\vec{r}\rho _{{\rm eq}}v^{2},\ \quad \ \ C_{LDM}=\int d\vec{r}\
\left( \frac{\delta ^{2}\varepsilon }{\delta \rho ^{2}}\right) _{{\rm eq}}%
\left[ \frac{\partial }{\partial r_{\nu }}(\rho _{{\rm eq}}v_{\nu })\right]
^{2},  \label{b}
\end{equation}
\begin{equation}
D_{0}=m\int d\vec{r}\rho _{{\rm eq}}v_{\alpha }v_{\nu }\frac{\partial
v_{\alpha }}{\partial r_{\nu }},\hspace{1cm}D_{1}=\int d\vec{r}\ \hat{%
\overline{L}}\left( P_{{\rm eq}}\overline{\Lambda }_{\alpha \nu }\right)
\frac{\partial v_{\alpha }}{\partial r_{\nu }},  \label{cld}
\end{equation}
\begin{equation}
D_{2}=\int d\vec{r}\ \frac{P_{{\rm eq}}^{2}}{\zeta }\overline{\Lambda }_{0}%
\overline{\Lambda }_{\alpha \nu }\frac{\partial v_{\alpha }}{\partial r_{\nu
}},\hspace{1cm}  \label{a1}
\end{equation}
\begin{equation}
\ A_{0}=\int d\vec{r}P_{{\rm eq}}\overline{\Lambda }_{\alpha \nu }\frac{%
\partial v_{\alpha }}{\partial r_{\nu }},\qquad A_{1}=m^{2}\int d\vec{r}\
\frac{P_{{\rm eq}}^{3}}{\xi }\overline{\Lambda }_{0}^{2}\overline{\Lambda }%
_{\alpha \nu }\frac{\partial v_{\alpha }}{\partial r_{\nu }},  \label{a01}
\end{equation}
\begin{equation}
A_{2}=m\int d\vec{r}\ \frac{P_{{\rm eq}}}{\zeta }\left( \overline{\Lambda }%
_{\alpha \nu }\hat{\overline{L}}(P_{{\rm eq}}\overline{\Lambda }_{0})+%
\overline{\Lambda }_{0}\hat{\overline{L}}(P_{{\rm eq}}\overline{\Lambda }%
_{\alpha \nu })\right) \frac{\partial v_{\alpha }}{\partial r_{\nu }},
\label{a2}
\end{equation}
\begin{equation}
A_{3}=2m^{2}\int d\vec{r}\ \frac{P_{{\rm eq}}^{3}}{\zeta ^{2}}\overline{%
\Lambda }_{0}^{2}\overline{\Lambda }_{\alpha \nu }\frac{\partial v_{\alpha }%
}{\partial r_{\nu }},\hspace{1cm}A_{4}=\int d\vec{r}\ \hat{\overline{L}}%
^{2}\left( P_{{\rm eq}}\overline{\Lambda }_{\alpha \nu }\right) \frac{%
\partial v_{\alpha }}{\partial r_{\nu }},  \label{a4}
\end{equation}
\begin{equation}
A_{5}=m\int d\vec{r}\ \hat{\overline{L}}\left( \frac{P_{{\rm eq}}^{2}}{\zeta
}\overline{\Lambda }_{0}\overline{\Lambda }_{\alpha \nu }\right) \frac{%
\partial v_{\alpha }}{\partial r_{\nu }},  \label{a5}
\end{equation}
with
\[
\hat{\overline{L}}P_{\alpha \beta }^{\prime }=v_{\nu }{\frac{\partial
P_{\alpha \beta }^{\prime }}{\partial r_{\nu }}}+P_{\alpha \beta }^{\prime }%
\frac{\partial v_{\nu }}{\partial r_{\nu }}+P_{\alpha \nu }^{\prime }{\frac{%
\partial v_{\beta }}{\partial r_{\nu }}}+P_{\beta \nu }^{\prime }{\frac{%
\partial v_{\alpha }}{\partial r_{\nu }}},
\]
\[
\overline{\Lambda }_{\alpha \beta }={\frac{\partial v_{\alpha }}{\partial
r_{\beta }}}+{\frac{\partial v_{\beta }}{\partial r_{\alpha }}}-{\frac{2}{3}}%
\delta _{\alpha \beta }\frac{\partial v_{\nu }}{\partial r_{\nu }},
\]
and
\[
\overline{\Lambda }_{0}=\frac{1}{2}(\overline{\Lambda }_{xx}+\overline{%
\Lambda }_{yy}-\overline{\Lambda }_{zz}).
\]
\ \ \ 

\end{document}